\documentclass[journal]{IEEEtran}
\usepackage{verbatim}
\usepackage{balance}
\usepackage{multirow}
\usepackage{booktabs}
\usepackage{amsfonts}
\usepackage{amsmath}
\usepackage{amsmath,cases}
\usepackage{amssymb}
\usepackage{graphicx}
\usepackage{cite}
\usepackage{epsfig}
\usepackage{url}
\usepackage{algorithm}
\usepackage{algorithmicx, algpseudocode}
\usepackage{epstopdf}
\usepackage{color}
\usepackage[tight]{subfigure}
\usepackage{mathrsfs}
\usepackage[justification=centering]{caption}
\usepackage{float}
\usepackage{diagbox}
\usepackage{tikz}
\usetikzlibrary{shapes,arrows,positioning}

\newcommand{\ds}{\displaystyle}

\newcommand{\la}{\langle}
\newcommand{\ra}{\rangle}

\newcommand{\bx}{\boldsymbol{x}}

\newcommand{\clC}{{\cal C}}
\newcommand{\clL}{{\cal L}}

\newcommand{\clR}{{\cal R}}

\newcommand{\bv}{\mathbf{v}}

\newcommand{\bV}{\mathbf{V}}

\newcommand{\ak}{a^{(\nu)}}

\newcommand{\bk}{b^{(\nu)}}

\newcommand{\gammak}{\lambda^{(\nu)}}

\newcommand{\ck}{c^{(\nu)}}

\newcommand{\clM}{{\cal M}}

\newcommand{\hke}{h^{e,(\nu)}}
\newcommand{\hka}{h^{a,(\nu)}}
\newcommand{\aka}{a^{a,(\nu)}}
\newcommand{\bka}{b^{a,(\nu)}}
\newcommand{\cka}{c^{a,(\nu)}}

\newcommand{\ake}{a^{e,(\nu)}}
\newcommand{\bke}{b^{e,(\nu)}}
\newcommand{\cke}{c^{e,(\nu)}}

\newcommand{\Qek}{\mathcal{A}^{e,(\nu)}}
\newcommand{\Qak}{\mathcal{A}^{a,(\nu)}}

\newcommand{\vka}{v^{a,(\nu)}}
\newcommand{\vkoa}{v^{a,(\nu+1)}}
\newcommand{\vke}{v^{e,(\nu)}}
\newcommand{\vkoe}{v^{e,(\nu+1)}}

\newcommand{\Psiak}{\Psi^{a,(\nu)}}
\newcommand{\Psiek}{\Psi^{e,(\nu)}}
\newcommand{\fak}{\gamma^{a,(\nu)}}
\newcommand{\fek}{\gamma^{e,(\nu)}}
\newcommand{\tclR}{\widetilde{\clR}}

\newcommand{\tPsika}{\widetilde{\Psi}^{a,(\nu)}}
\newcommand{\tPsike}{\widetilde{\Psi}^{e,(\nu)}}

\newcommand{\chinu}{\chi^{(\nu)}}
\allowdisplaybreaks
\begin{document}
\title{Long-Term Rate-Fairness-Aware  Beamforming Based Massive MIMO Systems}
\author{W. Zhu$^{1,2}$, H. D. Tuan$^2$,  E. Dutkiewicz$^2$, Y. Fang$^1$, H. V. Poor$^3$, and L. Hanzo$^4$
\thanks{The work was supported in part by the Australian Research Council's Discovery Projects under Grant DP190102501,  in part by the National Natural Science Foundation of China under Grant 61673253,  in part by the U.S National Science Foundation under Grants CNS-2128448 and ECCS-2335876, in part by the Engineering and Physical Sciences Research Council projects EP/W016605/1, EP/X01228X/1 and EP/Y026721/1 as well as of the European Research Council's Advanced Fellow Grant QuantCom (Grant No. 789028)}	
\thanks{$^1$School of Communication and Information Engineering, Shanghai University, Shanghai 200444, China
(email: wenbozhu@shu.edu.cn, yfang@staff.shu.edu.cn); $^2$School of Electrical and Data Engineering, University of Technology Sydney, Broadway, NSW 2007, Australia (email: wenbo.zhu@student.uts.edu.au, tuan.hoang@uts.edu.au, eryk.dutkiewicz@uts.edu.au); $^3$Department of Electrical and Computer Engineering, Princeton University, Princeton, NJ 08544, USA (email: poor@princeton.edu);
$^4$School of Electronics and Computer Science, University of Southampton, Southampton, SO17 1BJ, U.K (email: lh@ecs.soton.ac.uk) (Corresponding Author: Y. Fang) }
}
\date{}
\maketitle
\begin{abstract}
This is the first treatise on multi-user (MU)  beamforming designed for  achieving long-term rate-fairness in full-dimensional MU massive multi-input multi-output (m-MIMO) systems. Explicitly, based on the channel covariances, which can be assumed to be known  beforehand, we address this  problem by optimizing the following objective functions: the users' signal-to-leakage-noise ratios (SLNRs) using SLNR max-min optimization, geometric mean of SLNRs (GM-SLNR) based optimization, and SLNR soft max-min optimization. We develop a convex-solver based algorithm, which invokes a convex subproblem of cubic time-complexity  at each iteration for solving the  SLNR max-min  problem.  We then develop  closed-form expression based algorithms of scalable complexity for the solution of the GM-SLNR and of the SLNR soft max-min problem. The simulations provided confirm the users' improved-fairness ergodic rate distributions.
\end{abstract}
\section{Introduction}
Massive multi-input multi-output (m-MIMO)~\cite{MLYN16} schemes have
been shown to constitute a disruptive technology capable of providing
reliable quality of service for multiple users at a low transmission
power. With the explosive proliferation of transmit antennas, one has
to rely on a large two-dimensional (2D) uniformly spaced rectangular
antenna array (URA) to implement m-MIMO schemes
\cite{Nametal13,Kimetal14,Jietal17,Monetal15}. Transmit beamforming
(TBF) constitutes a prominent member of the m-MIMO family. The most
popular m-MIMO beamformers such as zero-forcing/regularized
zero-forcing and conjugate beamformers~\cite{MLYN16,NTDP19,Nguetal21}
rely on accurate channel state information (CSI), which is a
challenging task. Fortunately, the second-order statistics of
full-dimensional m-MIMO channels can be known
beforehand~\cite{S02,ANAC13} or efficiently signaled back to the
transmitter through feedback channels. In this context, it is of great
practical interest to consider the problem of designing statistical
beamformers that are capable of maintaining fairness in terms of the
users' ergodic rates, including their fair long-term
evolution. Unfortunately, this problem is computationally intractable,
because the ergodic rate function is not analytical in beamformers
\cite{BZGO10,Zhaetal17,LLJG18,QGCJ19,LYZ20,LLQJ20}. To resolve this
issue, statistical beamforming tends to rely on the so-called
signal-to-leakage-noise-ratio (SLNR) optimization~\cite{STS07}.  Due
to the statistical independence of the individual signal and its
leakage plus noise, the mathematical expectation of an individual SLNR
admits an analytical lower bound as the ratio of two convex quadratic
functions of the individual beamformer, thanks to the Mullen
theorem~\cite{M67}. With respect to the individual beamformer power, obtaining the optimal beamformer that maximizes this ratio can be readily achieved, thanks to the application of the Rayleigh-Ritz quotient theorem. This approach has been explored in previous studies using equal or fixed power allocation techniques
\cite{STS07,ZH11,PLSY13,KLS15,LJGH16,Zhaetal17,Sonetal19}. However, the equal-power allocation falls short of achieving rate-fairness, even in the case of zero-forcing or regularized zero-forcing beamforming ~\cite{NTDP19,Nguetal21}. Therefore, the problem of SLNR max-min optimization under a specific power constraint, which represents the most intuitive approach to SLNR optimization, still lacks a  computational solution.

\tikzset{
	block/.style = {rectangle, rounded corners, minimum width=2cm, minimum height=0.6cm, text centered, draw=black},
	arrow/.style = {-,>=stealth},
	every text node part/.style={align=center},
	execute at begin node=\setlength{\baselineskip}{12pt}
}
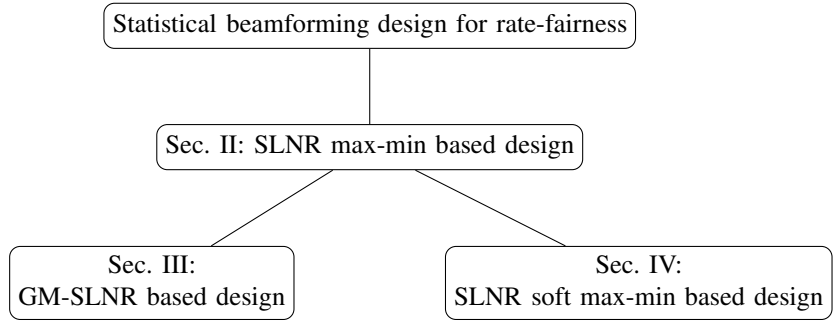
\begin{figure*}[!t]
	\centering
	\begin{tikzpicture}
		\node [block] (overview) {Statistical beamforming design for rate-fairness};
		\node [block, below = 1cm of overview] (sec2) {Sec. II: SLNR max-min based design};
		\coordinate [below = 1.5cm of sec2] (c);
		\node [block, left = of c] (sec3) {Sec. III: \\ GM-SLNR based design};
		\node [block, right = of c] (sec4) {Sec. IV: \\ SLNR soft max-min based design};
		\draw [arrow] (overview) -- (sec2);
		\draw [arrow] (sec2) -- (sec3);
		\draw [arrow] (sec2) -- (sec4);
	\end{tikzpicture}
	\caption{The outline of the paper highlighting the proposed beamforming designs.}
	\label{fig:outline}
\end{figure*}

Against the above background, this paper is the first one optimizing
the SLNRs for improving all users' ergodic rates. Explicitly, the
paper offers the following contributions:
\begin{itemize}
\item We develop an algorithm to
  tackle the aforementioned  max-min SLNR optimization problem. This algorithm
   iterates by solving convex problems (CPs);
\item Since there is no precise proportionality relationship between
  the SLNRs and the individual users' ergodic rates, we have to find a
  work-around. To elaborate, given that the SLNR max-min and ergodic
  rate max-min problems are not equivalent, we propose other SLNR
  optimization problems, such as maximizing the geometric mean of the
  SLNRs (GM-SLNR maximization) and a soft min SLNR (SLNR soft max-min)
  problem. We demonstrate that these objective functions achieve an
  improved minimum ergodic rate. Importantly, we devise algorithms that iterate by evaluating closed-form (CF) expressions of scalable complexity, rendering them practical solutions for m-MIMO systems;
\item The simulations provided reveal a surprising fact: the SLNR
  max-min optimization having the most balanced SLNR distributions
  associated with the highest minimum SLNR actually tends to result in
  the most unbalanced ergodic rate distributions. Unexpectedly, it also
  has the lowest minimum ergodic rate. By contrast, the soft max-min
  optimization exhibits the most balanced  ergodic rate
  distributions and has the highest minimum ergodic rate. {\em Thus, the
  soft max-min optimization has the best performance vs. complexity
  trade-off};
\item Statistical beamforming is based on channel statistics alone, hence
  it is free from channel-estimation overhead. As such, a byproduct of
  our results is a new beamforming design, which requires very limited
  feedback overhead. Hence it is eminently suitable
  for long-term multicast beamforming~\cite{L07} or MIMO-OFDM~\cite{L18}.
\end{itemize}
The paper is organized as follows. Section II is devoted to our
beamforming design relying on maximizing the minimum SLNR. By
contrast, the beamforming designs based on maximizing the GM-SLNR and
the soft min SLNR is considered in Section III and Section IV,
respectively. Our simulations are provided in Section V, while our
conclusions are discussed in Section VI. The Appendix provides
mathematical ingredients for the paper, which includes a new quadratic
minorant of a ratio of two quadratic functions. To help the reader to
smoothly follow the technical content of the paper, we provide the
flowchart in Fig. \ref{fig:outline}.

{\it Notation.} Boldface fonts are exclusively used to represent decision variables; The $N\times N$ identity matrix is represented by $I_N$; ${\cal C}(0,a)$ represents the set of circular Gaussian random variables having zero means and a variance of $a$; The dot product of matrices $X$ and $Y$ is given by $\la X,Y\ra=\mbox{trace}(X^HY)$, but we also use $\la X\ra$ to represent $\mbox{trace}(X)$; For the matrix $X$, the outer product $XX^H$ is denoted by  $[X]^2$; The expression $X\succeq 0$ indicates that the matrix $X$
is positive semi-definite; $\otimes$ represents the Kronecker product operator; Lastly,
\[
{\sf vec}\begin{bmatrix}x_1&\dots&x_L\end{bmatrix}\triangleq \begin{bmatrix}x_1\cr
\dots\cr
x_L\end{bmatrix}.
\]
Table \ref{notab} presents a summary of the used notations.
\begin{table*}[!t]
	\centering
	\caption{Notations used in the paper}
	\begin{tabular}{|c|l|}
		\hline
		Notation & \multicolumn{1}{c|}{Description}    \\
		\hline\hline
		$Q$&number of vertical and horizontal antennas at the base station \\ \hline
		$M$&number of outer products of the azimuth and elevation beamformers \\ \hline
		$L$&number of downlink users (DUs) \\ \hline
		$\bv_\ell^a,\bv_\ell^e$&azimuth and elevation beamformers for user $\ell$\\ \hline
		$\bv^a,\bv^e$&sets of azimuth and elevation beamformers\\ \hline
		$\gamma_\ell(\bv_\ell^a,\bv_\ell^e)$& SLNR for user $\ell$\\ \hline
		$\rho_{\ell}(\bv^a,\bv^e)$& achievable ergodic rate for user $\ell$\\ \hline
	\end{tabular}
\label{notab}
\end{table*}

\section{SLNR max-min based beamforming}
We consider a network supported by a base station (BS) equipped with a $Q\times Q$-uniform rectangular array (URA)
to serve $L$ single-antenna downlink users (DUs), which are indexed by $\ell\in\clL\triangleq \{1,\dots, L\}$.
The matrix $H_\ell\in\mathbb{C}^{Q\times Q}$ of entries $H_\ell(q,q')$
represents the channel spanning from the $(q,q')$-th antenna to DU $\ell$ is modeled by \cite{Yinetal14,NKA19}
\begin{equation}\label{mod1}
	H_\ell=\sqrt{\beta_\ell}\sqrt{R_{v,\ell}}H_{s,\ell}\sqrt{R_{h,\ell}},
\end{equation}
where  $\beta_\ell$ is a deterministic positive number representing the path-loss, $R_{v,\ell}$ and $R_{h,\ell}$ are deterministic positive definite matrices representing
the vertical and horizontal correlation matrices, while $H_{s,\ell}\in\mathbb{C}^{Q\times Q}$ is a random  matrix having
zero-mean and unit-variance identically distributed (iid) complex Gaussian elements, representing the small-scale
fading. Thus, we have
\begin{equation}\label{mod2}
	h_\ell\triangleq {\sf vec}(H_\ell)=\sqrt{\beta_\ell}(\sqrt{R_{h,\ell}}^T\otimes \sqrt{R_{v,\ell}} ){\sf vec}(H_{s,\ell}),
\end{equation}	
so
\begin{equation}\label{mod3}
	\mathbb{E}(h_\ell h_\ell^H)=\beta_\ell R_{h,\ell}^T\otimes R_{v,\ell}.
	\end{equation}
Furthermore,  $h_\ell^T=\sqrt{\beta_\ell}{\sf vec}^T(H_{s,\ell})(\sqrt{R_{h,\ell}}\otimes \sqrt{R_{v,\ell}} ^T)$, hence we have
\begin{equation}\label{mod4}
	\clR_\ell\triangleq 	\mathbb{E}[(h_\ell^T)^Hh_\ell^T]=\beta_\ell R_{h,\ell}\otimes R_{v,\ell}^T.
	\end{equation}
Additionally,
\begin{equation}\label{mod5}
	H_\ell^T=\sqrt{\beta_\ell}\sqrt{R_{h,\ell}}^TH^T_{s,\ell}\sqrt{R_{v,\ell}}^T,
\end{equation}
therefore
\begin{equation}\label{mod6}
	\tilde{h}_\ell\triangleq {\sf vec}(H_\ell^T)=\sqrt{\beta_\ell}(\sqrt{R_{v,\ell}}\otimes \sqrt{R_{h,\ell}}^T ){\sf vec}(H_{s,\ell}),
\end{equation}	
and $	\tilde{h}_\ell^T=\sqrt{\beta_\ell}{\sf vec}^T(H_{s,\ell})(\sqrt{R_{v,\ell}}^T\otimes \sqrt{R_{h,\ell}})$, for which we have
\begin{equation}\label{mod7}
	\tclR_\ell\triangleq 	\mathbb{E}[(\tilde{h}_\ell^T)^H\tilde{h}_\ell^T]=\beta_\ell R_{v,\ell}^T\otimes R_{h,\ell}.
\end{equation}	
Let $s_\ell\in \clC(0,1)$ be the information symbol intended for DU $\ell\in\clL$, which is  beamformed by the matrix $\bV_\ell\in \mathbb{C}^{Q\times Q}$ for the BS's downlink (DL) transmission, i.e. the $(q,q')$-th antenna transmits the signal $\sum_{\ell\in\clL}\bV_\ell(q,q')s_\ell$. The DL signal received at DU $\ell$ is given by
\begin{eqnarray}
	y_\ell&=&\sum_{\ell'\in\clL}\la H_\ell^T\bV_{\ell'}\ra s_{\ell'}+n_\ell,\label{td6e}
\end{eqnarray}
where $n_\ell\in\clC(0,\sigma)$ is the background noise at DU $\ell$.
 In the previous  paper \cite{Zhuetal23}, we proposed the following new class of $\bV_\ell\in\mathbb{C}^{Q\times Q}$
\begin{equation}\label{bea1}
\bV_\ell=\sum_{m\in\clM}\bv^e_{m,\ell}(\bv^a_{m,\ell})^T,
\end{equation}
with $\clM\triangleq \{1,\dots, M\}$.
In (\ref{bea1}), $\bv^e_{m,\ell}\in\mathbb{C}^{Q}$ plays the role of elevation beamforming, while $\bv^a_{m,\ell}\in\mathbb{C}^{Q}$ plays the role of azimuth beamforming.
The  $2D$-structured beamformer constructed as the outer product of the elevation and
azimuth beamformers \cite{Yinetal14,ALH17,Kanetal17tvt,Wanetal17,Sonetal19}
corresponds to  $M=1$, while the unstructured FD $\bV_\ell\in\mathbb{C}^{M\times M}$ corresponds to
 $M=Q$.  Importantly, it has been shown in \cite{Zhuetal23} that the beamformer (\ref{bea1}) constructed
for $M=2$ already achieves instantaneous rates (under the full availability of channel state information) close to those achieved by relying on $M=Q$. By contrast, the distinct objective of the present paper is to investigate its performance in terms of the ergodic rate achieved.

With $\bV_\ell$ defined by (\ref{bea1}), equation (\ref{td6e}) of the DL signal received at DU $\ell$ becomes
\begin{equation}\label{tra2}
y_\ell=\sum_{\ell'\in\clL}\left(\sum_{m\in\clM}(\bv^e_{m,\ell'})^TH_\ell\bv^a_{m,\ell'}\right)s_{\ell'}+n_\ell.
\end{equation}
In what follows, we will use the notations
\begin{equation}\label{td7a}
\bv^{z}_{\ell}\triangleq {\sf vec}
\begin{bmatrix}\bv^{z}_{1,\ell}&\dots&\bv^{z}_{M,\ell}\end{bmatrix}\in \mathbb{C}^{QM}, z\in\{a,e\},
\end{equation}
and then $\bv^a\triangleq \{\bv^a_{\ell}, \ell\in\clL \}$, and $\bv^e\triangleq \{\bv^e_{\ell}, \ell\in\clL \}$, and
$\bv\triangleq \{\bv^a,\bv^e\}$.
The achievable ergodic rate at DU $\ell$ is defined by
\begin{equation}\label{tra3}
\rho_\ell(\bv^a,\bv^e)\triangleq \mathbb{E}_{h_\ell}\left[ \ln\left(1+ {\sf SINR}_\ell(\bv^a,\bv^e) \right)\right],
\end{equation}
where we have
\begin{equation}\label{sinr1}
	{\sf SINR}_\ell(\bv^a,\bv^e)\triangleq \frac{|\sum_{m\in\clM}(\bv_{m,\ell}^e)^TH_\ell\bv^a_{m,\ell}|^2}{\sum_{\ell'\in\clL\setminus\{\ell\} }|\sum_{m\in\clM}(\bv_{m,\ell'}^e)^T H_\ell\bv^a_{m,\ell'}|^2+\sigma}.
\end{equation}
Given $(\bv^a,\bv^e)$, the statistical expectation in (\ref{tra3}) is calculated 
over all realization of $H_{\ell}$ with the covariance defined from (\ref{mod4}). This is contrasted 
to the instantaneous rate of $\ln\left(1+ {\sf SINR}_\ell(\bv^a,\bv^e) \right)$, which is calculated
at the specific instantaneous value of $H_{\ell}$.

Thus, the problem of maximizing the minimum ergodic rate is formulated as
\begin{subequations}\label{tra4}
	\begin{eqnarray}
		\max_{\bv=(\bv^a,\bv^e)}\  \min_{\ell\in\clL}\rho_{\ell}(\bv^a,\bv^e)\label{tra4a}\\
		\mbox{s.t.}\quad \sum_{\ell\in\clL}||\sum_{m\in\clM}\bv^a_{m,\ell}(\bv^e_{m,\ell})^T||^2\leq P.\label{tra4b}
	\end{eqnarray}
\end{subequations}
However, the ergodic rate function  $\rho_{\ell}$ is not computationally tractable and one has to use the so-called
SLNR defined by
\begin{equation}\label{slnr1}
\mathbb{E}_{h}\left[ \frac{ |\sum_{m\in\clM}(\bv_{m,\ell}^e)^TH_\ell\bv^a_{m,\ell}|^2}{\sum_{\ell'\in\clL\setminus\{\ell\} }|\sum_{m\in\clM}(\bv_{m,\ell}^e)^T H_{\ell'}\bv^a_{m,\ell}|^2+\sigma }  \right],	
\end{equation}	
and use the Mullen theorem \cite{M67} of $\mathbb{E}[x/y]\geq \mathbb{E}[x]/\mathbb{E}[y]$ for independent
random variables to obtain its lower bound
\begin{equation}\label{slnr2}
\gamma_\ell(\bv^a_\ell,\bv^e_\ell)\triangleq \frac{\mathbb{E}_{h_\ell}[ |\sum_{m\in\clM}(\bv_{m,\ell}^e)^TH_\ell\bv^a_{m,\ell}|^2]}
{\sum_{\ell'\in\clL\setminus\{\ell\} }\mathbb{E}_{h_{\ell'}}[|\sum_{m\in\clM}(\bv_{m,\ell}^e)^T H_{\ell'}\bv^a_{m,\ell}|^2]+\sigma}.
\end{equation}	
Thus, the most natural computationally tractable surrogate of problem (\ref{tra4}) representing a computationally intractable max-min ergodic rate optimization is the following problem
\begin{equation}\label{slnr3}
\max_{\bv^a,\bv^e}\gamma(\bv^a,\bv^e)\triangleq \min_{\ell\in\clL}\gamma_\ell(\bv_\ell^a,\bv_\ell^e)\quad \mbox{s.t.}\quad (\ref{tra4b}).	
\end{equation}
For convenience of terminology, we will still refer to this problem as SLNR max-min optimization.
This section aims to solve this problem.

Using the initialization of  $(v^{a,(0)},v^{e,(0)})$ that is feasible for (\ref{tra4b}), for $\nu=1, \dots$, and
letting $(\vka,\vke)$ be
a  feasible point of (\ref{tra4b}) that is found from the $(\nu-1)$-st iteration, we
process  the $\nu$-th iteration to generate $(\vkoa,\vkoe)$.
\subsection{Azimuth  beamforming optimization}
For $\bv^e$ fixed at $\vke$, the power constraint (\ref{tra4b})  is expressed by
\begin{equation}
\sum_{\ell\in\clL} \la \Qek_\ell,[\bv^a_\ell]^2\ra\leq P,\label{taa5b}
\end{equation}
with
\begin{align}\label{mek}
0\preceq \Qek_\ell&\triangleq \begin{bmatrix}\la \vke_{m,\ell},\vke_{m',\ell}\ra I_{M}\end{bmatrix}_{(m,m')\in \clM\times \clM}\nonumber\\ &\in\mathbb{C}^{(QM)\times (QM)}.
\end{align}
Let us now define the random variables
\begin{align}
\hke_{\ell',\ell}&\triangleq \begin{bmatrix}(\vke_{1,\ell'})^TH_\ell\dots&(\vke_{M,\ell'})^TH_\ell\end{bmatrix}\nonumber\\
&=\begin{bmatrix}H_\ell^T\vke_{1,\ell'}\cr
	\dots\cr
	H_\ell^T\vke_{M,\ell'}\end{bmatrix}^T\nonumber\\
&=\begin{bmatrix}\left((\vke_{1,\ell'})^T\otimes I_Q\right){\sf vec}(H_\ell^T)\cr
	\dots\cr
	\left((\vke_{M,\ell'})^T\otimes I_Q\right){\sf vec}(H_\ell^T)\end{bmatrix}^T\nonumber\\
&={\sf vec}^T(H_\ell^T)\begin{bmatrix}  \vke_{1,\ell'}\otimes I_Q&
	\dots&
	\vke_{M,\ell'}\otimes I_Q \end{bmatrix}\nonumber\\
&=\tilde{h}^T_\ell\Psiek_{\ell'} \in\mathbb{C}^{1\times (QM)},   (\ell',\ell)\in \clL\times\clL,
\label{sta1}
\end{align}
with $\tilde{h}_\ell$ defined from (\ref{mod6}),
and
\begin{equation}\label{sta3}
\Psiek_{\ell'}\triangleq \begin{bmatrix}  \vke_{1,\ell'}&
	\dots&
	\vke_{M,\ell'} \end{bmatrix}\otimes I_Q.
\end{equation}	
We then express $\gamma_\ell(\bv^a_\ell,\vke_\ell)$ by
\begin{align}
\gamma_\ell(\bv^a_\ell,\vke_\ell)&=\gamma^a_\ell(\bv^a_\ell)\nonumber\\
&\triangleq\frac{\mathbb{E}[|\hke_{\ell,\ell}
\bv^a_{\ell}|^2]}{\sum_{\ell'\in\clL\setminus\{ \ell\}}\mathbb{E}[|\hke_{\ell,\ell'}\bv^a_{\ell}|^2]+\sigma} \nonumber\\
&=\frac{||\sqrt{\tclR_\ell}\Psiek_\ell\bv^a_{\ell}||^2}{\sum_{\ell'\in\clL\setminus\{ \ell\}}||\sqrt{\tclR_{\ell'}}\Psiek_\ell\bv^a_{\ell}||^2+\sigma},\label{sta4}
\end{align}
where $\tclR_\ell$ is defined from (\ref{mod7}).
Thus, alternating optimization in $\bv^a$ is formulated by  the following problem:
\begin{equation}\label{taa5}
\max_{\bv^a} \min_{\ell\in\clL} \gamma^a_\ell(\bv^a_\ell)\quad
\mbox{s.t.}\quad (\ref{taa5b}).
\end{equation}
The following tight minorant of $\gamma^a_\ell(\bv^a_\ell)$ at $\vka$ is obtained by applying the inequality (\ref{ivt4}) for
 $(\bar{\chi},\bar{x})=(\bar{\chi}_\ell,\bar{x}_\ell)\triangleq
 ( \sqrt{\tclR_\ell}\Psiek_\ell\vka_{\ell}, \sum_{\ell'\in\clL\setminus\{ \ell\}}||\sqrt{\tclR_{\ell'}}\Psiek_\ell\vka_{\ell}||^2+\sigma  )$,
\begin{align}
\fak_\ell(\bv^a_\ell)\triangleq& \aka_\ell+2\Re\{\bka_\ell\bv^a_\ell\}\nonumber\\
&-\cka_\ell\sum_{\ell'\in \clL}||\sqrt{\tclR_{\ell'}}\Psiek_\ell\bv^a_{\ell}||^2,\label{taa7}
\end{align}
for
\begin{equation}\label{taa8}
\begin{array}{c}
\ds\aka_\ell\triangleq -\frac{||\bar{\chi}_{\ell}||^4}{\bar{x}_{\ell}^2}-\sigma \cka_\ell,\\
\bka_\ell\triangleq  \frac{\bar{x}_{\ell}+
||\bar{\chi}_{\ell}||^2}{\bar{x}_{\ell}^2}(\vka_{\ell})^H(\Psiek_\ell)^H\tclR_\ell\Psiek_\ell,\\
\ds\cka_\ell\triangleq \frac{||\bar{\chi}_{\ell}||^2}{\bar{x}_{\ell}^2}.
\end{array}
\end{equation}
Thus, we  solve the following convex  problem
\begin{equation}\label{taa12m}
\max_{\bv^a}\min_{\ell\in\clL}\fak_\ell(\bv^a_\ell)\quad\mbox{s.t.}\quad
(\ref{taa5b}),
\end{equation}
to generate $\vkoa$ satisfying
\begin{equation}\label{taa1}
\gamma(\vkoa,\vke)	>  \gamma(\vka,\vke)
\end{equation}
whenever  $ \gamma(\vka,\vke)\neq \gamma(\vkoa,\vke)$.
\subsection{Elevation beamforming optimization}
For  $\bv^a$  fixed at $\vkoa$, the power constraint (\ref{tra4b})  is formulated as
\begin{equation}
\sum_{\ell\in\clL} \la \Qak_\ell,[\bv^e_{\ell}]^2\ra\leq P,\label{tea5b}
\end{equation}
with
\begin{align}\label{mak}
\Qak_\ell&\triangleq \begin{bmatrix}\la \vkoa_{m,\ell},\vkoa_{m',\ell}\ra I_{M}
\end{bmatrix}_{(m,m')\in\clM\times \clM}\nonumber\\ &\in\mathbb{C}^{(QM)\times (QM)}.
\end{align}
Let us now define the random variables
\begin{align}
\hka_{\ell',\ell}&\triangleq \begin{bmatrix}(\vkoa_{1,\ell'})^TH_\ell^T&\dots&
	(\vkoa_{M,\ell'})^TH_\ell^T\end{bmatrix}\nonumber\\
&=\begin{bmatrix}H_\ell\vkoa_{1,\ell'}\cr
	\dots\cr
	H_\ell\vkoa_{M,\ell'}\end{bmatrix}^T\nonumber\\
&=\begin{bmatrix}\left((\vkoa_{1,\ell'})^T\otimes I_Q\right){\sf vec}(H_\ell)\cr
	\dots\cr
	\left((\vkoa_{M,\ell'})^T\otimes I_Q\right){\sf vec}(H_\ell)\end{bmatrix}^T\nonumber\\
&={\sf vec}^T(H_\ell)\begin{bmatrix}  \vkoa_{1,\ell'}\otimes I_Q&
	\dots&
	\vkoa_{M,\ell'}\otimes I_Q \end{bmatrix}\nonumber\\
&=h^T_\ell\Psiak_{\ell'}\in \mathbb{C}^{1\times (QM)},\label{ste1}
\end{align}
with $h_\ell$ defined from (\ref{mod2}),
and
\begin{equation}\label{ste3}
\Psiak_{\ell'}\triangleq \begin{bmatrix}  \vkoa_{1,\ell'}&
	\dots&
	\vkoa_{M,\ell'} \end{bmatrix}\otimes I_Q.
\end{equation}	
We then express $\gamma_\ell(\vkoa,\bv^e_\ell)$ by
\begin{align}
\gamma_\ell(\vkoa,\bv^e_\ell)&=\gamma^e_\ell(\bv^e_\ell)\nonumber\\
&\triangleq\frac{\mathbb{E}[|\hka_{\ell,\ell}\bv^e_{\ell}|^2]}{\sum_{\ell'\in\clL\setminus\{ \ell\}}\mathbb{E}[|\hka_{\ell,\ell'}\bv^e_{\ell}|^2]+\sigma} \nonumber\\
&=\frac{||\sqrt{\clR_\ell}\Psiak_\ell\bv^e_{\ell}||^2}{\sum_{\ell'\in\clL\setminus\{ \ell\}}||\sqrt{\clR_{\ell'}}\Psiak_\ell\bv^e_{\ell}||^2+\sigma},\label{ste4}
\end{align}
with $\clR_\ell$ defined from (\ref{mod4}).

The alternating optimization in $\bv^e$ is formulated by the problem:
 \begin{equation}\label{tae5}
 	\max_{\bv^e} \min_{\ell\in\clL} \gamma^e_\ell(\bv^e_\ell)\quad
 	\mbox{s.t.}\quad (\ref{tea5b}).
 \end{equation}
The following tight minorant of $f^e_\ell(\bv^e_\ell)$ at $\vke_\ell$ is obtained by  applying the inequality (\ref{ivt4})
 for $(\bar{\chi},\bar{x})=(\bar{\chi}_{\ell},\bar{x}_{\ell})\triangleq ( \sqrt{\clR_\ell}\Psiak_\ell\vke_{\ell}, \sum_{\ell'\in\clL\setminus\{ \ell\}}||\sqrt{\clR_{\ell'}}\Psiak_\ell\vke_{\ell}||^2 +\sigma )$,
 \begin{align}
 	\fek_\ell(\bv^e_\ell)\triangleq& \ake_\ell+2\Re\{\bke_\ell\bv^e_\ell\}\nonumber\\
 	&-\cke_\ell\sum_{\ell'\in \clL}||\sqrt{\clR_{\ell'}}\Psiak_\ell\bv^e_{\ell}||^2,\label{tea7}
 \end{align}
 for
 \begin{equation}\label{tea8}
 \begin{array}{c}
 \ds\ake_\ell\triangleq -\frac{||\bar{\chi}_{\ell}||^4}{\bar{x}_{\ell}^2}-\sigma \cke_\ell,\\
 \bke_\ell\triangleq  \frac{\bar{x}_{\ell}+||\bar{\chi}||^2}{\bar{x}_{\ell}^2}(\vke_{\ell})^H(\Psiak_\ell)^H\clR_\ell\Psiak_\ell,\\
 \ds\cke_\ell\triangleq \frac{||\bar{\chi}_{\ell}||^2}{\bar{x}_{\ell}^2}.\end{array}
 \end{equation}
 We now solve the following convex problem
 \begin{equation}\label{tea12m}
 	\max_{\bv^e}\min_{\ell\in\clL}\fek_\ell(\bv^e_\ell)\quad\mbox{s.t.}\quad
 	(\ref{tea5b})
 \end{equation}
to generate  $\vkoe$ satisfying
\begin{equation}\label{tae1}
	\gamma(\vkoa,\vkoe)>\gamma(\vkoa,\vke),
\end{equation}
whenever  $\gamma(\vkoa,\vkoe)\neq \gamma(\vkoa,\vke)$.
\begin{algorithm}[!t]
	\caption{SLNR max-min based statistical beamforming algorithm} \label{alg1}
	\begin{algorithmic}[1]
		\State \textbf{Initialization:} Initialize a feasible $(v^{a,(0)}, v^{e,(0)})$ for (\ref{tra4b}). Set $(v^{a,opt},
		v^{e,opt})=(v^{a,(0)}, v^{e,(0)})$ and $\mbox{rate}_{\min}=\min_{\ell\in\clL}\rho_\ell(v^{a,opt},
		v^{e,opt})$.
		\State \textbf{Repeat until convergence:} Generate
		$\vkoa$ by solving the convex problem  (\ref{taa12m}). If $\min_{\ell\in\clL} \rho_\ell(\vkoa,\vke)>\mbox{rate}_{\min}$,
		update $(v^{a,opt}, v^{e,opt})\leftarrow (\vkoa,\vke)$ and $\mbox{rate}_{\min}  \leftarrow \min_{\ell\in\clL} \rho_\ell(\vkoa,\vke)$.
		Generate   $\vkoe$ by solving the convex problem (\ref{tea12m}).  If $\min_{\ell\in\clL} \rho_\ell(\vkoa,\vkoe)>\mbox{rate}_{\min}$,
		update $(v^{a,opt}, v^{e,opt})\leftarrow (\vkoa,\vkoe)$ and $\mbox{rate}_{\min}  \leftarrow \min_{\ell\in\clL} \rho_\ell(\vkoa,\vkoe)$. Reset $\nu\leftarrow \nu+1$.
		\State \textbf{Output} $(v^{a,opt}, v^{e,opt})$ and $\mbox{rate}_{\min}$.
	\end{algorithmic}
\end{algorithm}
\subsection{Algorithm and its computational complexity}
Algorithm \ref{alg1} provides the pseudo code for computing (\ref{slnr3}) by solving the convex problems
(\ref{taa12m}) and (\ref{tea12m}) having the complexity order of ${\cal O} (Q^3M^3) $
to generate the next iterative feasible point  $(\vkoa,\vkoe)$, which according
to (\ref{taa1}) and (\ref{tae1}) satisfies $\gamma(\vkoa,\vkoe)>\gamma(\vka,\vke)$. The sequence $\{(\vka,\vke)\}$ thus
cannot be cyclic and thus it is convergent. Since having a higher minimum SLNR does not necessarily imply having a higher
minimum ergodic rate defined from (\ref{tra3})\footnote{Given the beamformers, the ergodic rate function
defined by (\ref{tra3}) can be calculated based on \cite{BZGO10}.},
 whenever generating a new feasible point for (\ref{slnr3}),
Algorithm \ref{alg1} also updates the incumbent beamformers achieving the best minimum user rate.

\section{GM-SLNR maximization based beamforming}
Note that the SLNR max min problem (\ref{slnr3}) is equivalent to the problem
\begin{eqnarray}\label{gme}
\max_{\bv^a,\bv^e}\min_{\ell\in\clL} \ln\left[1+\gamma_\ell(\bv_\ell^a,\bv_\ell^e)\right]\quad	\mbox{s.t.}\quad (\ref{tra4b}),
\end{eqnarray}
which  is seen computationally as difficult as problem (\ref{slnr3}). However, following our previous
developments \cite{Yuetaltwc22,Tuaetal22,Zhuetal22tvt}, it can be efficiently handled  via the following surrogate GM-SLNR problem:
\begin{equation}\label{gm}
\max_{\bv^a,\bv^e}\left[\prod_{\ell\in\clL} \ln\left(1+\gamma_\ell(\bv_\ell^a,\bv_\ell^e)\right) \right]^{1/L}	\quad\mbox{s.t.}\quad (\ref{tra4b}),
\end{equation}
i.e. maximizing the smooth objective function in (\ref{gm}) enhances the nonsmooth objective in (\ref{gme}).
The aim of this section is to provide a scalable algorithm for its computation. Initializing  $(v^{a,(0)},v^{e,(0)})$ that is feasible for (\ref{tra4b}), for $\nu=1, \dots$, we
let $(\vka,\vke)$ be a  feasible point of (\ref{tra4b}) that is found from the $(\nu-1)$-st iteration.
Similarly to \cite{Zhuetal22tvt,Zhuetal23}, the $\nu$-th iteration to generate $(\vkoa,\vkoe)$ is based on the following problem:
\begin{equation}\label{gmk}
\max_{\bv^a,\bv^e}\Gamma^{(\kappa)}(\bv^a,\bv^e)\triangleq \sum_{\ell\in\clL}\gammak_\ell\ln\left[1+\gamma_\ell(\bv^a_\ell,\bv^e_\ell)\right]\ \mbox{s.t.}\
(\ref{tra4b}),
\end{equation}
for\footnote{$\left[\prod_{\ell\in\clL} \ln\left(1+\gamma_\ell(\bv_\ell^a,\bv_\ell^e)\right) \right]^{1/L}=\min_{\lambda_\ell>0, \prod_{\ell=1}^L\lambda_\ell = 1}\\ \left[\sum_{\ell=1}^{L}\lambda_\ell \ln\left[1+\gamma_\ell(\bv_\ell^a,\bv_\ell^e)\right] \right]$, which is achieved at\\
$\lambda_\ell= \frac{\max_{\ell'\in\clL}\ln\left[1+\gamma_{\ell'}(\vka,\vke)\right]}{\ln\left[1+\gamma_\ell(\vka,\vke)\right]}, \ell\in\clL$. }
\begin{equation}\label{gammak}
\gammak_\ell\triangleq  \frac{\max_{\ell'\in\clL}\ln\left[1+\gamma_{\ell'}(\vka,\vke)\right]}{\ln\left[1+\gamma_\ell(\vka,\vke)\right]}, \ell\in\clL.
\end{equation}
\subsection{Azimuth beamforming optimization}
The alternating optimization on $\bv^a$ is based on optimizing $\bv^a$ with $\bv^e$ fixed at $\vke$, which is
based on the following problem
\begin{equation}\label{gmk1}
\max_{\bv^a}\Gamma^{(\kappa)}(\bv^a,\vke)=\sum_{\ell\in\clL}\gammak_\ell \ln\left[1+\gamma^a_\ell(\bv^a_\ell)\right]\
\mbox{s.t.}\ (\ref{taa5b}),
\end{equation}
with $\gamma^a_\ell(\bv^a_\ell)$ defined from (\ref{sta4}).

The following minorant of $\Gamma^{(\kappa)}(\bv^a,\vke)$ at $\vka$ is obtained by applying the inequality (\ref{inv2}) for $(\bar{\chi}, \bar{x})=(\bar{\chi}^a_{\ell}, \bar{x}^a_{\ell}) \triangleq \left(\sqrt{\tclR_\ell}\Psiek_\ell\vka_{\ell}, \sum_{\ell'\in\clL\setminus\{ \ell\}}||\sqrt{\tclR_{\ell'}}\Psiek_\ell\vka_{\ell}||^2+\sigma\right)$:
\begin{align}
\Gamma^{a,(\kappa)}(\bv^a)\triangleq&\sum_{\ell\in\clL}\gammak_\ell\left(\aka_\ell
+2\Re\{\bka_\ell\bv^a_\ell\}\right.\nonumber\\
&\left.-\cka_\ell\sum_{\ell'\in \clL}||\sqrt{\tclR_{\ell'}}\Psiek_\ell\bv^a_{\ell}||^2 \right),\label{gmk2}
\end{align}
where
\begin{equation}\label{gmk3}
\begin{array}{c}
\aka_\ell\triangleq \alpha(\bar{\chi}^a_{\ell}, \bar{x}^a_{\ell}),\ \bka_\ell\triangleq \frac{(\bar{\chi}^a_\ell)^H\sqrt{\tclR_\ell}\Psiek_\ell }{\bar{x}^a_\ell},\\
\cka_\ell\triangleq \psi(\bar{\chi}^a_{\ell}, \bar{x}^a_{\ell}),\\
\alpha(\bar{\chi}^a_{\ell}, \bar{x}^a_{\ell})\triangleq
	\ln\left(1+\frac{||\bar{\chi}^a_{\ell}||^2}{\bar{x}^a_{\ell}}\right)-
	\frac{||\bar{\chi}^a_{\ell}||^2}{\bar{x}^a_{\ell}}.
\end{array}
\end{equation}

We thus solve the following problem to generate $\vkoa$:
\begin{equation}\label{gmk3}
\max_{\bv^a}\Gamma^{a,(\kappa)}(\bv^a)\quad
\mbox{s.t.}\quad (\ref{taa5b}).
\end{equation}
The optimal solution of this problem is given by the
following closed-form under the definition of $\tPsika_\ell\triangleq \cka_\ell(\Psiek_\ell)^H
\left(\sum_{\ell'\in\clL}\tclR_{\ell'}\right)\Psiek_\ell$,
\begin{equation}\label{gmk4}
\vkoa_\ell=\begin{cases}\begin{array}{l}(\tPsika_\ell)^{-1}(\bka_\ell)^H\\ \mbox{if}\ \sum_{\ell\in\clL}
\la \Qek_\ell,[(\tPsika_\ell)^{-1}(\bka_\ell)^H]^2\ra\leq P,\cr
\gammak_\ell\left(\gammak_\ell\tPsika_\ell+\mu I_{QM}\right)^{-1}(\bka_\ell)^H\\ \mbox{otherwise},
\end{array}
\end{cases}
\end{equation}
where $\mu>0$ is found by bisection so that
\[
\sum_{\ell\in\clL}\la \Qek_\ell,[\gammak_\ell\left(\gammak_\ell\tPsika_\ell+\mu I_{QM}\right)^{-1}(\bka_\ell)^H]^2\ra= P.
\]
\subsection{Elevation beamforming optimization}
The alternating optimization on $\bv^e$ is based on optimizing $\bv^e$ with $\bv^a$ fixed at $\vkoa$, which is based on the following problem
\begin{equation}\label{egmk1}
\max_{\bv^a}\Gamma^{(\kappa)}(\vkoa,\bv^e)=\sum_{\ell\in\clL}\gammak_\ell \ln\left[1+\gamma^e_\ell(\bv^e_\ell)\right]\
\mbox{s.t.}\ (\ref{tea5b}),
\end{equation}
with $\gamma^e_\ell(\bv^e_\ell)$ defined by (\ref{ste4}).

The following minorant of $\Gamma^{(\kappa)}(\vkoa,\bv^e)$ at $\vke$ is obtained by
applying the inequality (\ref{inv2}) for $(\bar{\chi},\bar{x})=(\bar{\chi}^e_\ell,\bar{x}^e_\ell)\triangleq
\left(\sqrt{\clR_\ell}\Psiak_\ell\vke_{\ell}, \sum_{\ell'\in\clL\setminus\{ \ell\}}||\sqrt{\clR_{\ell'}}\Psiak_\ell\vke_{\ell}||^2+\sigma\right)$,
\begin{align}
\Gamma^{e,(\kappa)}(\bv^e)\triangleq&\sum_{\ell\in\clL}\gammak_\ell\left(\ake_\ell+
2\Re\{\bke_\ell\bv^e_\ell\}\right.\nonumber\\
&\left.-\cke_\ell\sum_{\ell'\in \clL}||\sqrt{\clR_{\ell'}}\Psiak_\ell\bv^e_{\ell}||^2 \right),\label{egmk2}
\end{align}
where
\begin{equation}\label{egmk3}
\begin{array}{c}
\ake_\ell\triangleq \alpha(\bar{\chi}^e_\ell,\bar{x}^e_\ell),\ \bke_\ell\triangleq \frac{(\bar{\chi}^e_\ell)^H\sqrt{\clR_\ell}\Psiak_\ell }{\bar{x}^e_\ell},\\
\cke_\ell\triangleq \psi(\bar{\chi}^e_\ell,\bar{x}^e_\ell).
\end{array}
\end{equation}
We thus solve the following problem to generate $\vkoa$:
\begin{equation}\label{egmk3}
\max_{\bv^e}\Gamma^{e,(\kappa)}(\bv^e)\quad
\mbox{s.t.}\quad (\ref{tea5b}).
\end{equation}
The optimal solution of this problem is given by  the following
closed-form under the definition of  $\tPsike_\ell\triangleq \cke_\ell(\Psiak_\ell)^H
\left(\sum_{\ell'\in\clL}\clR_{\ell'}\right)\Psiak_\ell$,
\begin{equation}\label{egmk4}
\vkoe_\ell=\begin{cases}\begin{array}{l}(\tPsike_\ell)^{-1}(\bke_\ell)^H\\ \mbox{if}\ \sum_{\ell\in\clL}
\la \Qak_\ell,[(\tPsike_\ell)^{-1}(\bke_\ell)^H]^2\ra\leq P,\cr
\gammak_\ell\left(\gammak_\ell\tPsike_\ell+\mu I_{QM}\right)^{-1}(\bke_\ell)^H\\ \mbox{otherwise},
\end{array}
\end{cases}
\end{equation}
where $\mu>0$ is found by bisection so that
\[
\sum_{\ell\in\clL}\la \Qak_\ell,[\gammak_\ell\left(\gammak_\ell\tPsike_\ell+\mu I_{QM}\right)^{-1}(\bke_\ell)^H]^2\ra= P.
\]
\subsection{Scalable algorithm and its computational complexity}
Algorithm \ref{alggm} provides the pseudo-code of computing (\ref{gm}) and also for updating the beamformers achieving the best minimum ergodic rate. The computational complexity of each iteration is on the order of ${\cal O} (QM)$.
Its convergence can be shown by
following the proof provided in \cite{Zhuetal23,Zhuetal22tvt}.
\begin{algorithm}[!t]
	\caption{GM-SLNR based statistical beamforming algorithm} \label{alggm}
	\begin{algorithmic}[1]
	\State \textbf{Initialization:} Initialize a feasible  $(v^{a,(0)}, v^{e,(0)})$  for (\ref{tra4b}).  Set $(v^{a,opt},
	v^{e,opt})=(v^{a,(0)}, v^{e,(0)})$ and $\mbox{rate}_{\min}=\min_{\ell\in\clL}\rho_\ell(v^{a,opt},
	v^{e,opt})$.
		\State \textbf{Repeat until convergence:} Generate
$\vkoa$ by  the closed form (\ref{gmk4}).  If $\min_{\ell\in\clL} \rho_\ell(\vkoa,\vke)>\mbox{rate}_{\min}$,
update $(v^{a,opt}, v^{e,opt})\leftarrow (\vkoa,\vke)$ and $\mbox{rate}_{\min}  \leftarrow \min_{\ell\in\clL} \rho_\ell(\vkoa,\vke)$.
Generate   $\vkoe$ by the closed form  (\ref{egmk4}).  If $\min_{\ell\in\clL} \rho_\ell(\vkoa,\vkoe)>\mbox{rate}_{\min}$,
update $(v^{a,opt}, v^{e,opt})\leftarrow (\vkoa,\vkoe)$ and $\mbox{rate}_{\min}  \leftarrow \min_{\ell\in\clL} \rho_\ell(\vkoa,\vkoe)$.
Reset $\nu\leftarrow \nu+1$.
 \State \textbf{Output}  $(v^{a,opt}, v^{e,opt})$ and $\mbox{rate}_{\min}$.
\end{algorithmic}
\end{algorithm}

\section{SLNR soft max-min optimization based beamforming}
The SLNR max min problem (\ref{slnr3}) is also equivalent to the problem
\begin{eqnarray}\label{soft}
	\max_{\bv^a,\bv^e}\min_{\ell\in\clL} \ln\left(1+\frac{\gamma_\ell(\bv_\ell^a,\bv_\ell^e)}{c}\right)\quad	\mbox{s.t.}\quad (\ref{tra4b}),
\end{eqnarray}
for any $c>0$.  Then we have
\begin{equation}\label{soft1}
\min_{\ell\in\clL} \ln\left(1+\frac{\gamma_\ell(\bv_\ell^a,\bv_\ell^e)}{c}\right)=
-\max_{\ell\in\clL}\ln\left(1+\frac{\gamma_\ell(\bv_\ell^a,\bv_\ell^e)}{c}\right)^{-1}
\end{equation}	
while
\begin{align}
&\max_{\ell\in\clL}\ln\left(1+\frac{\gamma_\ell(\bv_\ell^a,\bv_\ell^e)}{c}\right)^{-1}\nonumber\\
\leq&\ln \frac{\sum_{\ell\in\clL} \left(1+\frac{\gamma_\ell(\bv_\ell^a,\bv_\ell^e)}{c}\right)^{-1}}{L}+\ln L\label{soft2}\\
\leq&\max_{\ell\in\clL}\ln\left(1+\frac{\gamma_\ell(\bv_\ell^a,\bv_\ell^e)}{c}\right)^{-1}+\ln L.\label{soft3}
\end{align}
For small $c$, $\ln L$ is small compared to the nonsmooth function in the LHS of (\ref{soft2}) and the RHS
of (\ref{soft3}), so  we have
\begin{align}
&\max_{\ell\in\clL}\ln\left(1+\frac{\gamma_\ell(\bv_\ell^a,\bv_\ell^e)}{c}\right)^{-1}\nonumber\\
\approx& 	\ln \frac{\sum_{\ell\in\clL} \left(1+\frac{\gamma_\ell(\bv_\ell^a,\bv_\ell^e)}{c}\right)^{-1}}{L}\label{soft4}\\
=&\ln \left(\sum_{\ell\in\clL} \left(1+\frac{\gamma_\ell(\bv_\ell^a,\bv_\ell^e)}{c}\right)^{-1}\right)-\ln L.\label{soft5}
\end{align}	
Therefore, we can approximate the problem (\ref{soft}) of nonsmooth function  optimization by the following problem of
smooth function optimization:
\begin{equation}\label{sm}
	\max_{\bv^a,\bv^e}\left[-  \ln\left(\sum_{\ell\in\clL}\left(1+\frac{\gamma_\ell(\bv_\ell^a,\bv_\ell^e)}{c}\right)^{-1}\right)+  \ln L \right]	\ \mbox{s.t.}\  (\ref{tra4b}),
\end{equation}
which is equivalent to the problem
\begin{equation}\label{sme}
	\min_{\bv^a,\bv^e}  \tilde{\Gamma}(\bv^a,\bv^e)\triangleq \ln\left(\sum_{\ell\in\clL}\left(1+\frac{\gamma_\ell(\bv_\ell^a,\bv_\ell^e)}{c}\right)^{-1}\right)	\ \mbox{s.t.}\ (\ref{tra4b}).	
\end{equation}
In this paper, we refer to (\ref{sm})/(\ref{sme}) as  SLNR soft max-min optimization. This section aims for developing an algorithm of scalable complexity for solving (\ref{sme}) and optimizing the users' ergodic rates.

Use the initialization  $(v^{a,(0)},v^{e,(0)})$ that is feasible for (\ref{tra4b}), for $\nu=1, \dots$,
and let $(\vka,\vke)$ be
a  feasible point of (\ref{tra4b}) that is found from the $(\nu-1)$-st iteration. Then we can carry out
the $\nu$-th iteration to generate $(\vkoa,\vkoe)$.
\subsection{Azimuth beamforming optimization}
With $\gamma^a_\ell(\bv^a_\ell)$ defined from (\ref{sta4}), the alternating optimization in $\bv^a$ is based on the problem
\begin{equation}\label{smk1}
\min_{\bv^a}\tilde{\Gamma}^a(\bv^a)\quad
\mbox{s.t.}\quad (\ref{taa5b}),
\end{equation}
where we have $\tilde{\Gamma}^a(\bv^a)$ given by (\ref{smk2}) shown at the top of next page.
\begin{figure*}[t]
\begin{align}
\tilde{\Gamma}^a(\bv^a)&\triangleq\tilde{\Gamma}(\bv^a,\vke)\nonumber\\
&=\ln\left(\sum_{\ell\in\clL}\left(1+\frac{\gamma^a_\ell(\bv_\ell^a)}{c}\right)^{-1}\right)\nonumber\\
&=\ln\left(\sum_{\ell\in\clL}\left(1-\frac{||\sqrt{\tclR_\ell}\Psiek_\ell\bv^a_{\ell}||^2}{||\sqrt{\tclR_\ell}\Psiek_\ell\bv^a_{\ell}||^2
+c\left(\sum_{\ell'\in\clL\setminus\{ \ell\}}||\sqrt{\tclR_{\ell'}}\Psiek_\ell\bv^a_{\ell}||^2+\sigma \right)}\right)\right).\label{smk2}
\end{align}
\hrulefill
\vspace{-0.2cm}
\end{figure*}

The following tight majorant of $\tilde{\Gamma}^a(\bv^a)$ at $\vka$ is obtained
by applying the inequality (\ref{ap6}) for $(\bar{\chi}_\ell,\bar{x}_\ell)=\left(\sqrt{\tclR_\ell}\Psiek_\ell\vka_{\ell},
\sum_{\ell'\in\clL\setminus\{ \ell\}}||\sqrt{\tclR_{\ell'}}\Psiek_\ell\vka_{\ell}||^2+\sigma\right)$:
\begin{align}
	\tilde{\Gamma}^{a,(\kappa)}(\bv^a)\triangleq& \ak_1-2\sum_{\ell\in\clL}\Re\{\bk_{1,\ell}\bv^a_\ell\}\nonumber\\
	&+\sum_{\ell\in\clL}\ck_{1,\ell}\left(||\sqrt{\tclR_\ell}\Psiek_\ell\bv^a_{\ell}||^2  \right.\nonumber\\
	&\left.\quad +c\sum_{\ell'\in\clL\setminus\{ \ell\}}||\sqrt{\tclR_{\ell'}}\Psiek_\ell\bv^a_{\ell}||^2\right)\label{smk3}\\
	=&\ak_1-2\sum_{\ell\in\clL}\Re\{\bk_{1,\ell}\bv^a_\ell\}+\sum_{\ell\in\clL}(\bv^a_{\ell})^H\tPsika_\ell\bv^a_\ell,\label{smk3a}
\end{align}
where we have
\begin{subequations}\label{smk4}
\begin{align}
	\ak_1&\triangleq \tilde{\Gamma}^a(\vka)\nonumber\\ &\quad + \frac{1}{\chinu_1}\sum_{\ell\in\clL}\left(\frac{||\bar{\chi}_\ell||^2}
	{c\bar{x}_\ell+||\bar{\chi}_\ell||^2} + \frac{||\bar{\chi}_\ell||^2}{(c\bar{x}_\ell+||\bar{\chi}_\ell||^2)^2}c\sigma\right),
	\label{smk4a}\\	
	\bk_{1,\ell}&\triangleq \frac{1}{\chinu_1}\frac{(\sqrt{\tclR_\ell}\Psiek_\ell\vka_{\ell})^H\sqrt{\tclR_\ell}\Psiek_\ell}{c\bar{x}_\ell
	+||\bar{\chi}_\ell||^2}, \ell\in\clL,
	\label{smk4b}\\
	\ck_{1,\ell}&\triangleq \frac{1}{\chinu_1}\frac{||\bar{\chi}_\ell||^2}{(c\bar{x}_\ell+||\bar{\chi}_\ell||^2)^2}, \ell\in\clL,\label{smk4c}\\
\chinu_1&\triangleq \sum_{\ell\in\clL}\left(1+\frac{\gamma^a_\ell(\vka_\ell)}{c}\right)^{-1},
\end{align}
\end{subequations}
and
\begin{align}\label{smk5}
\tPsika_\ell\triangleq& \ck_{1,\ell}\left( (\Psiek_\ell)^H\tclR_\ell\Psiek_\ell\right.\nonumber\\
&\left.+c\sum_{\ell'\in\clL\setminus\{ \ell\}}(\Psiek_\ell)^H\tclR_{\ell'}\Psiek_\ell \right).
\end{align}
We thus solve the following problem
\begin{equation}\label{smk6}
\min_{\bv^a}\tilde{\Gamma}^{a,(\kappa)}(\bv^a)\quad
\mbox{s.t.}\quad (\ref{taa5b}).
\end{equation}
to generate $\vkoa$ so that
\begin{equation}\label{softo1}
\tilde{\Gamma}(\vkoa,\vke)<\tilde{\Gamma}(\vka,\vke).
\end{equation}
The optimal solution of this problem is given by  the following closed-form:
\begin{equation}\label{gmk4}
\vkoa_\ell=\begin{cases}\begin{array}{ll}(\tPsika_\ell)^{-1}(\bka_\ell)^H\\\mbox{if}\ \sum_{\ell\in\clL}
\la \Qek_\ell,[(\tPsika_\ell)^{-1}(\bka_\ell)^H]^2\ra\leq P,\cr
\left(\tPsika_\ell+\mu I_{QM}\right)^{-1}(\bka_\ell)^H\\ \mbox{otherwise},
\end{array}
\end{cases}
\end{equation}
where $\mu>0$ is found by bisection so that $\sum_{\ell\in\clL}\la \Qek_\ell,[\left(\tPsika_\ell+\mu I_{QM}\right)^{-1}(\bka_\ell)^H]^2\ra= P$.
\subsection{Elevation beamforming optimization}
With $\gamma^e_\ell(\bv^e_\ell)$ defined by (\ref{ste4}),
the alternating optimization in $\bv^e$ is based on the problem
\begin{equation}\label{esmk1}
\min_{\bv^e}\tilde{\Gamma}^e(\bv^e)\quad
\mbox{s.t.}\quad (\ref{tea5b}),
\end{equation}
where $\tilde{\Gamma}^e(\bv^e)$ is given by (\ref{esmk2}) shown at the top of next page.
\begin{figure*}[t]
\begin{align}
\tilde{\Gamma}^e(\bv^e)&\triangleq\tilde{\Gamma}(\vkoa,\bv^e)\nonumber\\
&=\ln\left(\sum_{\ell\in\clL}\left(1+\frac{\gamma^e_\ell(\bv_\ell^e)}{c}\right)^{-1}\right)\nonumber\\
&=\ln\left(\sum_{\ell\in\clL}\left(1-\frac{||\sqrt{\clR_\ell}\Psiak_\ell\bv^e_{\ell}||^2}{
||\sqrt{\clR_\ell}\Psiak_\ell\bv^e_{\ell}||^2
+c\left(\sum_{\ell'\in\clL\setminus\{ \ell\}}||\sqrt{\clR_{\ell'}}\Psiak_\ell\bv^e_{\ell}||^2+\sigma \right)}\right)\right).\label{esmk2}
\end{align}
\hrulefill
\vspace{-0.2cm}
\end{figure*}

The following tight majorant of $\tilde{\Gamma}^e(\bv^e)$ at $\vka$ is obtained by applying the inequality (\ref{ap6}) for $(\bar{\chi}_\ell, \bar{x}_\ell)=\left(\sqrt{\clR_\ell}\Psiak_\ell\vke_{\ell},
\sum_{\ell'\in\clL\setminus\{ \ell\}}||\sqrt{\clR_{\ell'}}\Psiek_\ell\vke_{\ell}||^2+\sigma\right)$:
\begin{align}
	\tilde{\Gamma}^{e,(\kappa)}(\bv^e)\triangleq& \ak_2-2\sum_{\ell\in\clL}\Re\{\bk_{2,\ell}\bv^e_\ell\}\nonumber\\
	&+\sum_{\ell\in\clL}\ck_{2,\ell}\left(||\sqrt{\clR_\ell}\Psiak_\ell\bv^e_{\ell}||^2 \right.\nonumber\\
	&\left.\quad +c\sum_{\ell'\in\clL\setminus\{ \ell\}}||\sqrt{\clR_{\ell'}}\Psiak_\ell\bv^e_{\ell}||^2
	\right)\label{esmk3}\\
	=&\ak_2-2\sum_{\ell\in\clL}\Re\{\bk_{2,\ell}\bv^e_\ell\}+\sum_{\ell\in\clL}(\bv^e_{\ell})^H\tPsike_\ell\bv^e_\ell,\label{emk3a}
\end{align}
where we have
\begin{subequations}\label{esmk4}
\begin{align}
	\ak_2&\triangleq \tilde{\Gamma}^e(\vke)\nonumber\\ &\quad+ \frac{1}{\chinu_2}\sum_{\ell\in\clL}\left(\frac{||\bar{\chi}_\ell||^2}
	{c\bar{x}_\ell+||\bar{\chi}_\ell||^2} + \frac{||\bar{\chi}_\ell||^2}{(c\bar{x}_\ell+||\bar{\chi}_\ell||^2)^2}c\sigma\right),
	\label{esmk4a}\\	
	\bk_{2,\ell}&\triangleq \frac{1}{\chinu_2}\frac{(\sqrt{\clR_\ell}\Psiak_\ell\vke_{\ell})^H\sqrt{\clR_\ell}\Psiak_\ell}{c\bar{x}_\ell
	+||\bar{\chi}_\ell||^2}, \ell\in\clL,
	\label{esmk4b}\\
	\ck_{2,\ell}&\triangleq \frac{1}{\chinu_2}\frac{||\bar{\chi}_\ell||^2}{(c\bar{x}_\ell+||\bar{\chi}_\ell||^2)^2}, \ell\in\clL,\label{esmk4c}\\
	\chinu_2&\triangleq \sum_{\ell\in\clL}\left(1+\frac{\gamma^e_\ell(\vke_\ell)}{c}\right)^{-1},
\end{align}
\end{subequations}
and
\begin{align}\label{esmk5}
\tPsike_\ell\triangleq& \ck_{2,\ell}\left( (\Psiak_\ell)^H\clR_\ell\Psiak_\ell\right.\nonumber\\
&\left.+c\sum_{\ell'\in\clL\setminus\{ \ell\}}(\Psiak_\ell)^H\clR_{\ell'}\Psiak_\ell \right).
\end{align}
We thus solve the following problem
\begin{equation}\label{esmk6}
\min_{\bv^e}\tilde{\Gamma}^{e,(\kappa)}(\bv^e)\quad
\mbox{s.t.}\quad (\ref{tea5b}),
\end{equation}
to generate $\vkoe$ so that
\begin{equation}\label{softo2}
\tilde{\Gamma}(\vkoa,\vkoe)<\tilde{\Gamma}(\vkoa,\vke).
\end{equation}
The optimal solution of this problem is given by the following closed-form:
\begin{equation}\label{esmk6}
\vkoe_\ell=\begin{cases}\begin{array}{ll}(\tPsike_\ell)^{-1}(\bke_\ell)^H\\\mbox{if}\ \sum_{\ell\in\clL}
\la \Qak_\ell,[(\tPsike_\ell)^{-1}(\bke_\ell)^H]^2\ra\leq P,\cr
\left(\tPsike_\ell+\mu I_{QM}\right)^{-1}(\bke_\ell)^H\\\mbox{otherwise},
\end{array}
\end{cases}
\end{equation}
where $\mu>0$ is found by bisection so that $\sum_{\ell\in\clL}\la \Qak_\ell,[\left(\tPsike_\ell+\mu I_{QM}\right)^{-1}(\bke_\ell)^H]^2\ra= P$.
\subsection{Scalable algorithm and its computational complexity}
Algorithm \ref{algsm} provides the pseudo code for solving problem (\ref{sme}) by
generating the next iterative feasible point  $(\vkoa,\vkoe)$ by the closed forms (\ref{gmk4})
and (\ref{esmk6}), which according
to (\ref{softo1}) and (\ref{softo2}) satisfies $	\tilde{\Gamma}(\vkoa,\vkoe)<	\tilde{\Gamma}(\vka,\vke)$. The sequence $\{(\vka,\vke)\}$ thus
cannot be cyclic and thus it is convergent. Whenever generating a new feasible point for (\ref{sme}),
Algorithm \ref{algsm} also updates the incumbent beamformers achieving the best minimum user rate.
The computational complexity of each iteration is on the order of ${\cal O} (QM) $. Given the cubic-time computational complexity of the convex-solver based Algorithm 1, both Algorithm 2 and Algorithm 3 have scalable complexities, rendering them computationally efficient.

\begin{algorithm}[!t]
	\caption{SLNR Soft max-min based statistical beamforming  algorithm} \label{algsm}
	\begin{algorithmic}[1]
	\State \textbf{Initialization:} Initialize a feasible $(v^{a,(0)}, w^{e,(0)})$  for
(\ref{tea5b}). Set $(v^{a,opt},
	v^{e,opt})=(v^{a,(0)}, v^{e,(0)})$ and $\mbox{rate}_{\min}=\min_{\ell\in\clL}\rho_\ell(v^{a,opt},
	v^{e,opt})$.
		\State \textbf{Repeat until convergence of the objective function in (\ref{sme}):} Generate
$\vkoa$ by  (\ref{gmk4}). If $\min_{\ell\in\clL} \rho_\ell(\vkoa,\vke)>\mbox{rate}_{\min}$,
update $(v^{a,opt}, v^{e,opt})\leftarrow (\vkoa,\vke)$ and $\mbox{rate}_{\min}  \leftarrow \min_{\ell\in\clL} \rho_\ell(\vkoa,\vke)$.
Generate $\vkoe$ by (\ref{esmk6}). If $\min_{\ell\in\clL} \rho_\ell(\vkoa,\vkoe)>\mbox{rate}_{\min}$,
update $(v^{a,opt}, v^{e,opt})\leftarrow (\vkoa,\vkoe)$ and $\mbox{rate}_{\min}  \leftarrow \min_{\ell\in\clL} \rho_\ell(\vkoa,\vkoe)$. Reset $\nu\leftarrow \nu+1$.
 \State \textbf{Output} $(v^{a,opt}, v^{e,opt})$ and $\mbox{rate}_{\min}$.
	\end{algorithmic}
\end{algorithm}

\section{Simulation Results}
A $Q \times Q$-URA BS is positioned at the center of a cell having a $200$-meter radius. Within the cell, there are $L=10$ DUs that are randomly placed. The BS antennas have a height of $10$ meters, while the DUs have a height of $1.5$ meters.
The path-loss and shadow-fading of DU $\ell$ experienced at a distance $d_\ell$ from the BS is given by \cite{3GPP} $\rho_\ell = 28.8 + 35.3 \log10(d_\ell)+\xi_\ell$ (in dB), where $\xi_\ell\sim \mathcal{N}(0,\sigma_\mathrm{sf}^2)$ represents the shadow-fading coefficient with $\sigma_\mathrm{sf}=7.8$. The carrier frequency is set to $2$ GHz, while the bandwidth is set to $10$ MHz. The power density of the background noise is set to $-174$ dBm/Hz. A convergence tolerance of $10^{-3}$ is specified for the algorithms with respect to the SLNR performance. We refer to the $m$-th vertical and $n$-th horizontal direction of the URA as the $(m,n)$-th antenna element. In accordance with \cite{Yinetal14}, the vertical correlation matrix $R_{v,\ell}$ is represented by
\begin{equation}
	[R_{v,\ell}]_{m,p}=e^{j\pi(p-m)\cos\beta_\ell}e^{-\frac{1}{2}[\sigma_{\beta}\pi(p-m)\sin\beta_\ell]^2},
\end{equation}
while the horizontal correlation matrix is formulated as
\begin{equation}
	[R_{h,\ell}]_{n,q}=\frac{1}{\sqrt{\gamma_3}}e^{-\frac{\gamma_2^2\cos^2\alpha_\ell}{2\gamma_3}}e^{j\frac{\gamma_1\cos\alpha}{\gamma_3}}e^{-\frac{(\gamma_1\sigma_{\alpha}\sin\alpha_\ell)^2}{2\gamma_3}},
\end{equation}
where
$\gamma_1=\pi(q-n)\sin\beta_\ell$,
$\gamma_2=\sigma_{\beta}\pi(q-n)\cos\beta_\ell$,
$\gamma_3=\gamma_2^2\sigma_{\alpha}^2\sin^2\alpha_\ell+1$.
The azimuth and elevation angles of DU $\ell$, denoted as $\alpha_\ell$ and $\beta_\ell$, respectively, are determined based on the relative location of the DU with respect to the BS. Furthermore, $\sigma_{\alpha}$ and $\sigma_{\beta}$ denote the angular spreads in the azimuth and elevation domains, respectively, and both are set to $5^{\circ}$.

We will compare the minimum ergodic rate achieved by our optimized beamforming with the following baseline uniform power beamformers given by \cite{STS07}
\begin{equation}\label{bl1}
	\sqrt{P/K} w_{u,\ell},
\end{equation}	
where  $w_{u,\ell}\in\mathbb{C}^{Q^2}$ is the eigenvalue vector corresponding to
the largest eigenvalue of the matrix
\[
\left( \sum_{\ell'\in\clL\setminus\{\ell\}}\clR_{\ell'}+(K\sigma/P)I_{Q^2}  \right)^{-1}\clR_\ell
\]
with $\clR_\ell$ defined from (\ref{mod4}).

Once a better feasible point $(v^{a,(\nu+1)},v^{e,(\nu+1)})$ is obtained in the $\nu$-th iteration, which maximizes the minimum SLNR, or in its GM form or soft max-min form, we define $V^{(\nu+1)}_{\ell}$ by
(\ref{bea1}) and then $v^{(\nu+1)}_{\ell}\triangleq {\sf vec}(V^{(\nu+1)}_{\ell})$, and
$W^{(\nu+1)}\triangleq \begin{bmatrix}v^{(\nu+1)}_1&\dots&v^{(\nu+1)}_L\end{bmatrix}$. Furthermore, we
define $S_{\ell}=(W^{(\nu+1)})^H\clR_{\ell}W^{(\nu+1)}$ and then $\tilde{S}_{\ell}$ by removing the $\ell$-th column and the $\ell$-th row of $S_{\ell}$. The ergodic rate is calculated according to \cite[Th.1]{BZGO10} as follows:
\begin{align}
	\mathbb{E}\{\ln(1+\mathrm{SINR}_\ell)\}
	=&\sum_{m=1}^{\mathrm{rank}(S_\ell)} \frac{e^{\frac{\sigma}{\mu_m}}E_1(\frac{\sigma}{\mu_m})}
	{\prod_{l\ne m}(1-\frac{\mu_l}{\mu_m})}\nonumber\\
	&-\sum_{m=1}^{\mathrm{rank}(\tilde{S}_\ell)} \frac{e^{\frac{\sigma}{\lambda_m}}E_1(\frac{\sigma}{\lambda_m})}
	{\prod_{l\ne m}(1-\frac{\lambda_l}{\lambda_m})},
\end{align}
where $\mu_1,\dots,\mu_{\mathrm{rank(S_\ell)}}$ and $\lambda_1,\dots,\lambda_{\mathrm{rank(\tilde{S}_\ell)}}$ are the distinct non-zero eigenvalues of $S_\ell$ and $\tilde{S}_\ell$, respectively, while  $E_1(x)=\int_{1}^{\infty} e^{-xu}/udu$ is the exponential integral.

We use the following legends to specify the proposed implementations: ``MM1/MM2'' refer to the CP-based algorithms with $M=1/M=2$; ``G1/G2'' refer to the GM-based algorithms with $M=1/M=2$; ``Soft1/Soft2'' refer to the soft max-min based algorithms with $M=1/M=2$; ``EQ'' refers to the baseline algorithm that uses uniform power beamformers.

\begin{figure}[!t]
	\centering
	\includegraphics[width=0.9\linewidth]{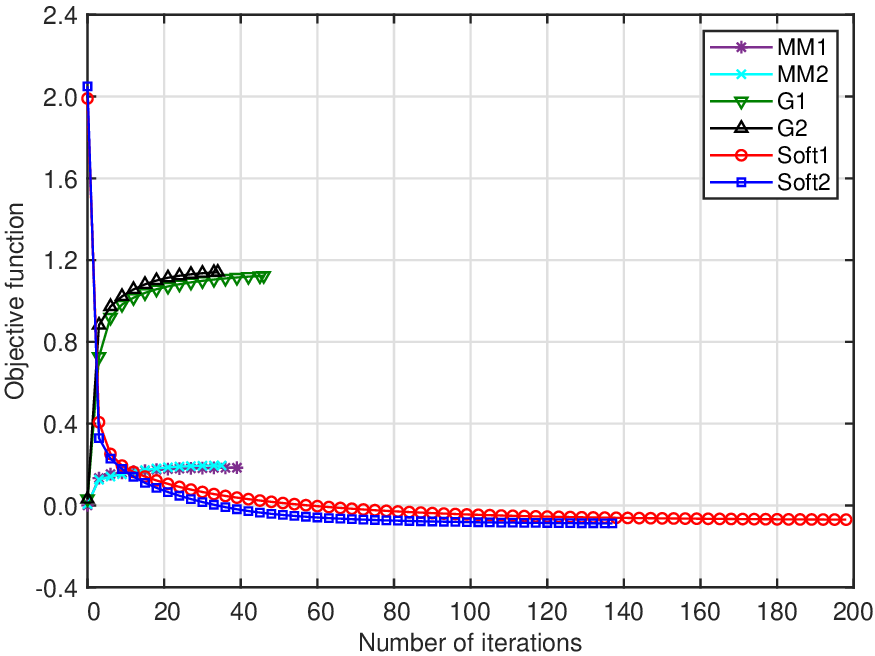}
	\caption{The convergence of the algorithms.}
	\label{fig:conv_obj_fun}
\end{figure}

\begin{figure}[!t]
	\centering
	\includegraphics[width=0.9\linewidth]{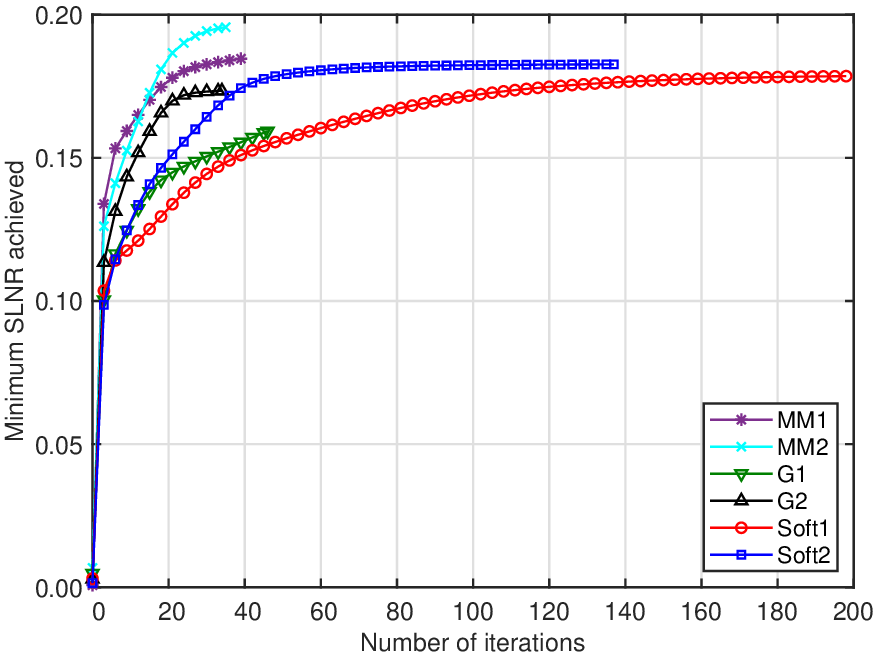}
	\caption{The convergence of the minimum SLNR.}
	\label{fig:conv_min_SLNR}
\end{figure}

\begin{figure}[!t]
	\centering
	\includegraphics[width=0.9\linewidth]{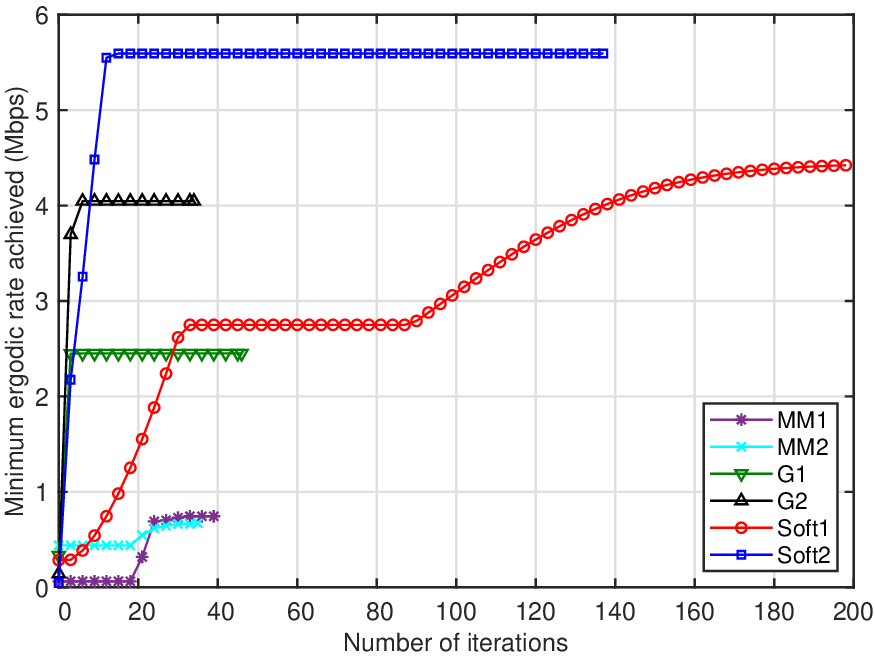}
	\caption{The convergence of the minimum ergodic rate.}
	\label{fig:conv_min_rate}
\end{figure}

\begin{table}[!t]
	\centering
	\caption{Minimum ergodic rate (Mbps) vs. coefficient $c$ for the soft max-min algorithm.}
	\begin{tabular}{|l|c|c|c|}
		\hline
		& $c=1$  & $c=0.1$ & $c=0.01$ \\ \hline
		Soft1    & 2.97 & 4.95  & 4.27   \\ \hline
		Soft2    & 2.82 & 5.34  & 5.26   \\ \hline
	\end{tabular}
	\label{table:soft_min_rate_vs_c}
\end{table}

Fig. \ref{fig:conv_obj_fun} depicts the convergence of the algorithms, while Fig. \ref{fig:conv_min_SLNR} and Fig. \ref{fig:conv_min_rate} portray the convergence of the minimum SLNR and the minimum ergodic rate achieved, respectively. The transmit power budget $P$ is maintained at $24$ dBm, and the number of BS antennas is kept fixed at 64. The coefficient $c$ used in soft max-min based algorithms is set to 0.1. It is worth noting that we use the minimization form (\ref{sme}) for the objective function of the soft max-min based algorithm. Therefore, in Fig. \ref{fig:conv_obj_fun}, the objective functions of Soft1/Soft2 show a gradually decreasing trend until convergence.
A noteworthy observation from Fig. \ref{fig:conv_min_SLNR} and Fig. \ref{fig:conv_min_rate} is that the minimum SLNR exhibits a gradual increase with each iteration, while the minimum ergodic rate does not follow this pattern. This implies that the beamformers that yield the highest SLNR values may not necessarily lead to the highest minimum ergodic rate.

Importantly, the choice of the coefficient $c$ in the soft max-min algorithm should be determined based on specific factors, such as the dimensions of the beamformers and the transmit power budget $P$. We present Table \ref{table:soft_min_rate_vs_c} to illustrate the influence of different $c$ values on the minimum ergodic rate achieved (in Mbps). The results are obtained at a transmit power $P$ of 24 dBm and 64 BS antennas. We can observe that setting $c$ to 0.1 yields the highest minimum ergodic rate.

\begin{figure}[!t]
	\centering
	\includegraphics[width=0.9\linewidth]{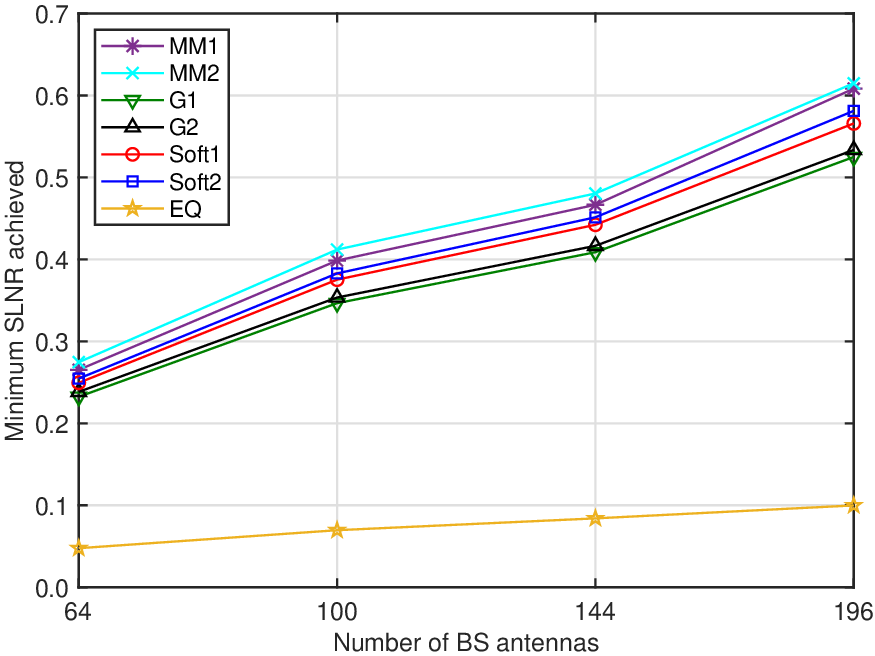}
	\caption{The minimum SLNR achieved vs. the number of BS antennas.}
	\label{fig:min_SLNR_vs_M}
\end{figure}

We evaluate the performance of the proposed algorithms upon varying the numbers of BS antennas. Fig. \ref{fig:min_SLNR_vs_M} plots the minimum SLNR achieved versus the number of BS antennas at a power of $P = 24$ dBm. The convex-solver based algorithms achieve the highest minimum SLNR, followed by the soft max-min based algorithms, and then the GM-based algorithms. The minimum SLNR obtained by the baseline algorithm, which calculates the minimum ergodic rate using the uniform power beamformers, is significantly lower than that of the other algorithms. Moreover, using the sum of two outer products yields better results than using the sum of just one outer product.

\begin{figure}[!t]
	\centering
	\includegraphics[width=0.9\linewidth]{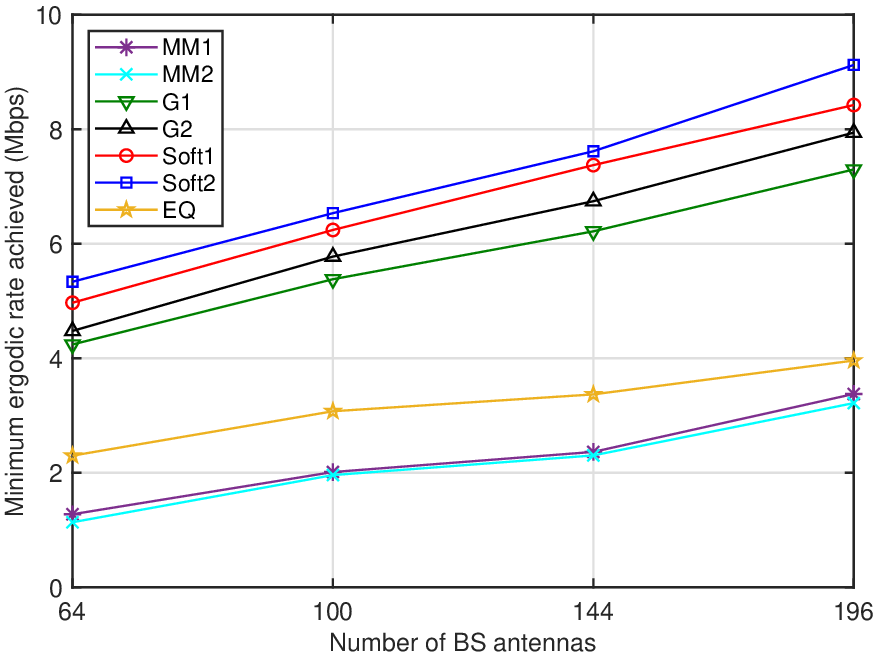}
	\caption{The minimum ergodic rate achieved vs. the number of BS antennas.}
	\label{fig:min_rate_vs_M}
\end{figure}

Fig. \ref{fig:min_rate_vs_M} shows the minimum ergodic rate achieved versus the number of BS antennas at the same power of $P = 24$ dBm. It is worth noting that the minimum ergodic rate is determined by selecting the maximum value obtained during the iterative optimization process, rather than the one calculated by the beamformers that yields the highest SLNR.
The results demonstrate that both the GM-based and soft max-min based algorithms outperform the baseline algorithm, which calculates the minimum ergodic rate using the uniform power beamformers. Additionally, the soft max-min based algorithms achieve a higher minimum ergodic rate than the  GM-based algorithms. We can also find that the sum of two outer products performs better than the sum of one outer product in terms of the ergodic rate. However, the minimum ergodic rates obtained by the convex-solver based algorithms are significantly lower than those obtained by the other algorithms, including the baseline algorithm. The baseline algorithm performs better than expected, probably due  to its connection with regularized zero forcing beamforming
under the equal power allocation \cite{PAD12,BJ13}.

\begin{figure}[!t]
	\centering
	\includegraphics[width=0.9\linewidth]{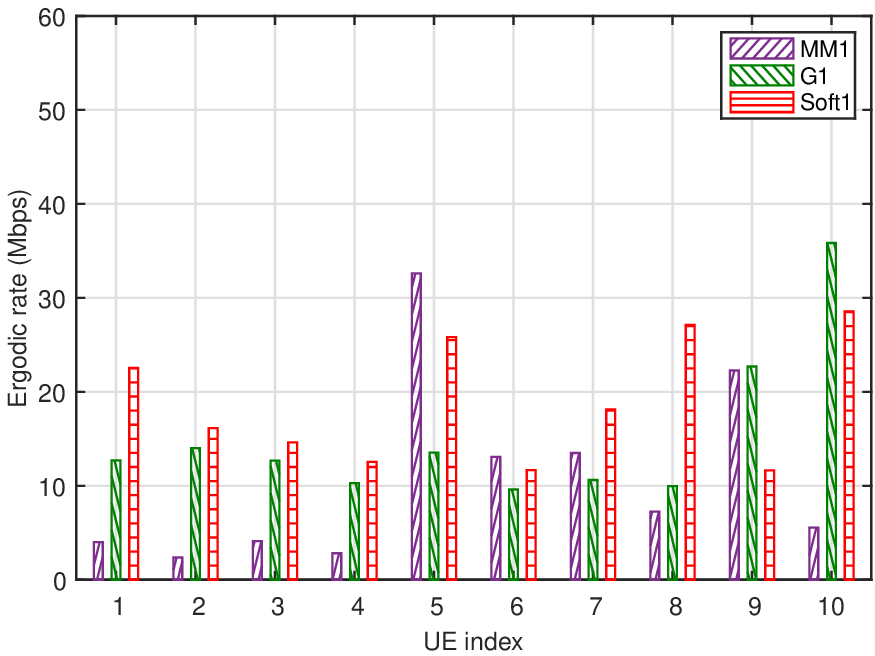}
	\caption{Ergodic rate distribution for three algorithms associated with $M= 1$.}
	\label{fig:user_rate_M14_Q1}
\end{figure}

\begin{figure}[!t]
	\centering
	\includegraphics[width=0.9\linewidth]{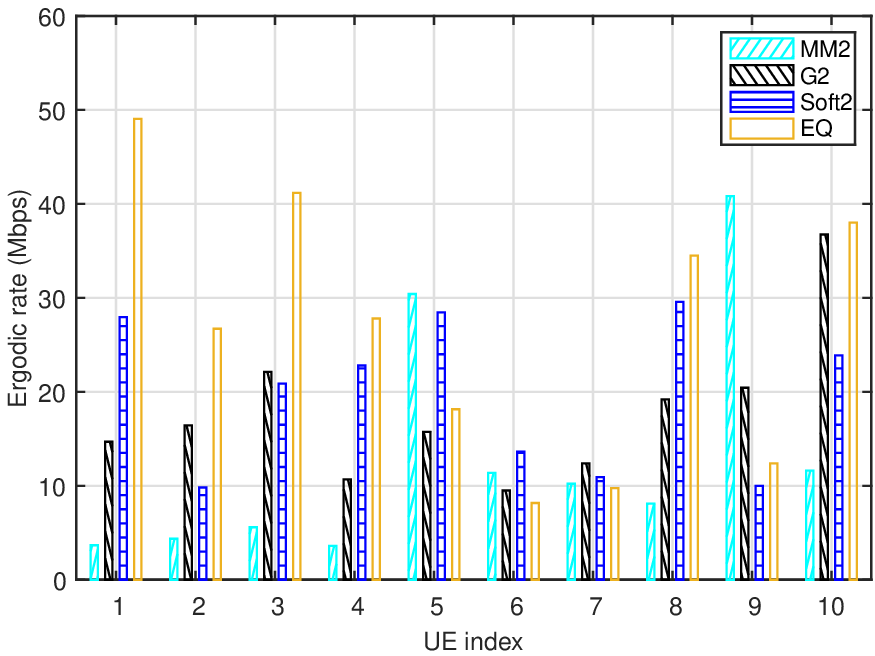}
	\caption{Ergodic rate distribution for three algorithms associated with $M = 2$ and the baseline algorithm.}
	\label{fig:user_rate_M14_Q2}
\end{figure}

Fig. \ref{fig:user_rate_M14_Q1} portrays the users' ergodic rate distribution pattern for algorithms associated with $M = 1$, while Fig. \ref{fig:user_rate_M14_Q2} portrays this pattern for algorithms associated with $M = 2$ and the baseline algorithm. This simulation uses 196 antennas at the BS and a power of $P = 24$ dBm. Similar to Fig. \ref{fig:min_rate_vs_M}, these figures
illustrate that both the GM-based and the soft max-min based algorithms G2 and Soft2 succeed in enhancing the ergodic rates of all DUs, while also ensuring their minimum value. Furthermore, the baseline algorithm EQ performs better than the SLNR max-min based M2, probably owing  to its association with regularized zero forcing beamforming
under equal power allocation \cite{PAD12,BJ13}.

\begin{table}[!t]
	\centering
	\caption{The minimum-to-maximum ergodic rate ratio vs. the number of BS antennas}
	\begin{tabular}{|l|c|c|c|c|}
		\hline
		& 64 antennas & 100 antennas & 144 antennas & 196 antennas \\ \hline
		MM1 & 0.0519 & 0.0760   & 0.0810   & 0.1014   \\ \hline
		MM2 & 0.0536 & 0.0780   & 0.0838   & 0.1039   \\ \hline
		G1 & 0.1106 & 0.1421   & 0.1510   & 0.1589   \\ \hline
		G2 & 0.1303 & 0.1494   & 0.1713   & 0.1759   \\ \hline
		Soft1 & 0.1410 & 0.1675   & 0.1859   & 0.1915   \\ \hline
		Soft2 & 0.1794 & 0.1861   & 0.2007   & 0.2264   \\ \hline
		EQ & 0.0281 & 0.0345   & 0.0355   & 0.0386   \\ \hline
	\end{tabular}
	\label{table:ratio_rate}
\end{table}

\begin{table}[!t]
	\centering
	\caption{Jain's fairness index of the ergodic rate vs. the number of BS antennas}
	\begin{tabular}{|l|c|c|c|c|}
		\hline
		& 64 antennas & 100 antennas & 144 antennas & 196 antennas \\ \hline
		MM1 & 0.4901 & 0.5528   & 0.5690   & 0.5849   \\ \hline
		MM2 & 0.4807 & 0.5613   & 0.5729   & 0.5833   \\ \hline
		G1 & 0.6303 & 0.6875   & 0.7087   & 0.7186   \\ \hline
		G2 & 0.6706 & 0.7007   & 0.7250   & 0.7261   \\ \hline
		Soft1 & 0.6942 & 0.7211   & 0.7371   & 0.7399   \\ \hline
		Soft2 & 0.7138 & 0.7286   & 0.7456   & 0.7635   \\ \hline
		EQ & 0.4930 & 0.5236   & 0.5478   & 0.5573   \\ \hline
	\end{tabular}
	\label{table:jain_rate}
\end{table}

For more qualitative analysis, Table \ref{table:ratio_rate} and Table \ref{table:jain_rate} present the minimum-to-maximum ergodic rate ratio and Jain's fairness index \cite{Jain84} of the ergodic rate, respectively, at $P = 24$ dBm. These two metrics are used for evaluating the distribution pattern of the ergodic rate obtained by the proposed algorithms.
As the number of BS antennas increases from 64 to 196, both the minimum-to-maximum ergodic rate ratio and Jain's fairness index increase for our proposed algorithms, which illustrates that employing more antennas for beamforming flexibility helps achieve more uniformly distributed ergodic user rates.
Moreover, the soft max-min based algorithms yield the highest minimum-to-maximum ergodic rate ratio and Jain's fairness index, demonstrating their ability to achieve a more fair ergodic rate distribution. Additionally, both the GM-based and soft max-min based algorithms outperform the convex-solver based algorithms and the baseline algorithm in terms of rate fairness.

\begin{figure}[!t]
	\centering
	\includegraphics[width=0.9\linewidth]{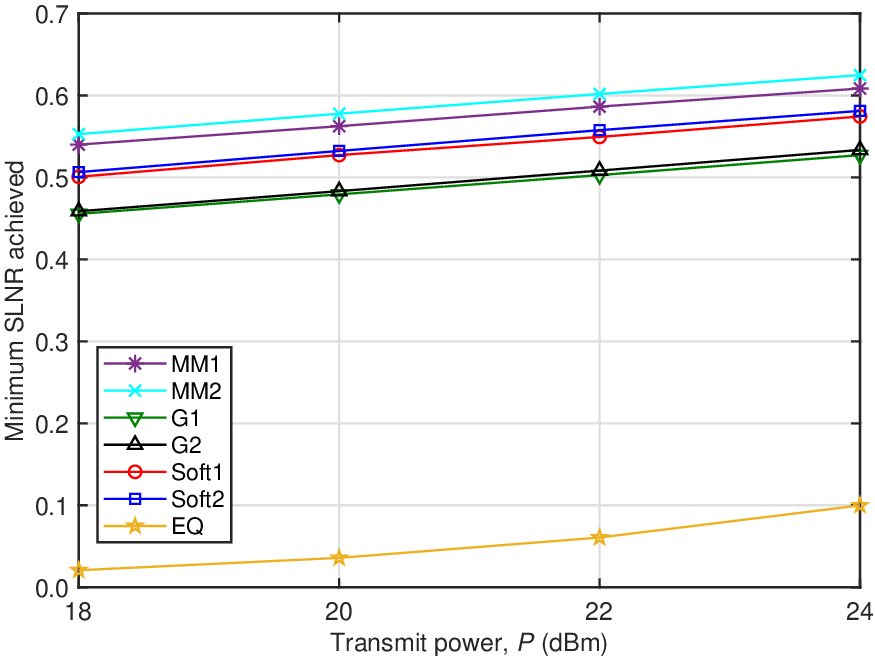}
	\caption{The minimum SLNR achieved vs. $P$ using 196 antennas at the BS.}
	\label{fig:min_SLNR_vs_P}
\end{figure}

\begin{figure}[!t]
	\centering
	\includegraphics[width=0.9\linewidth]{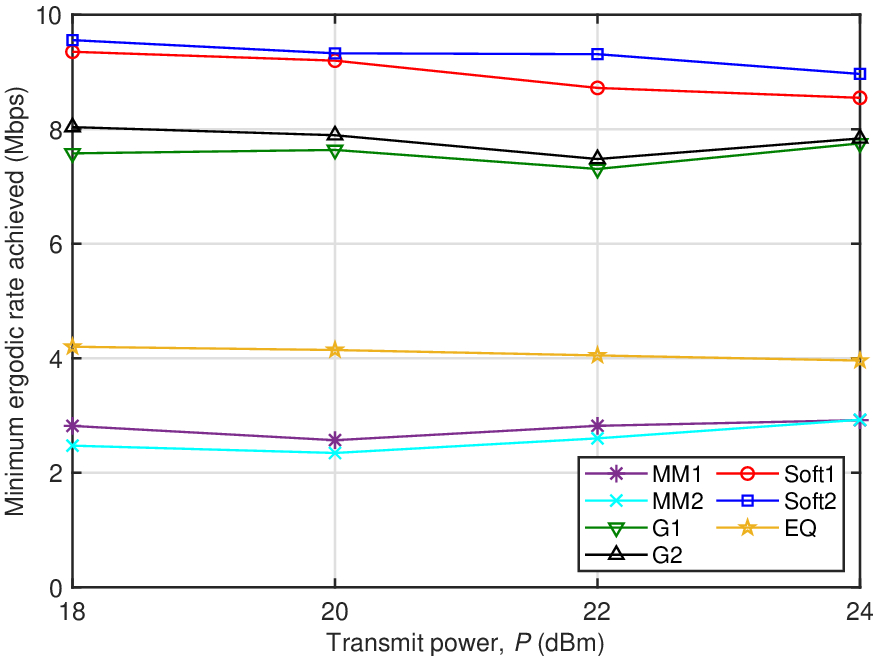}
	\caption{The minimum ergodic rate achieved vs. $P$ using 196 antennas at the BS.}
	\label{fig:min_rate_vs_P}
\end{figure}

In addition, we investigate the effect of the transmit power budget $P$ on the minimum SLNR and minimum ergodic rate achieved, as depicted in
Fig. \ref{fig:min_SLNR_vs_P} and Fig. \ref{fig:min_rate_vs_P}, for a fixed number of 196 BS antennas. As previous noted in Fig. \ref{fig:conv_min_rate}, the highest SLNR value does not necessarily guarantee the highest minimum ergodic rate. Observe that the minimum SLNR increases with $P$, but the minimum ergodic rate does not exhibit the same trend. However, similar to the observations gleaned from Fig. \ref{fig:min_rate_vs_M}, the soft max-min based and GM-based algorithms perform better than the baseline algorithm, while the convex-solver based algorithms yield the lowest minimum ergodic rate.

\section{Conclusions}
In order to investigate the long-term ergodic rate distributions provided by full-dimensional m-MIMO systems, we have proposed different statistical beamforming schemes, which are based on the optimization of the users' SLNRs.
Accordingly, the paper has developed computationally  efficient algorithms for their
solutions, which are also used for generating feasible beamforming in updating the incumbent ergodic minimum rate. Furthermore, we have provided simulations to show the impact of SLNR metric on the ergodic rates. An extension to statistical hybrid beamforming for mmWave communications is currently under our further investigation.
\section*{Appendix: mathematical ingredients}
The following inequalities hold true for all $\pmb{\chi}\in\mathbb{C}^{N}$, $\bx>0$ and $\bar{\chi}\in\mathbb{C}^N$, $\bar{x}>0$:
\begin{equation}\label{inv2}
	\ln\left(1+\frac{||\pmb{\chi}||^2}{\bx}\right)\geq \alpha(\bar{\chi},\bar{x})+
	2\frac{\Re\{\bar{\chi}^H\pmb{\chi}\}}{\bar{x}}
	-\psi(\bar{\chi},\bar{x})(||\pmb{\chi}||^2+\bx),
\end{equation}
with
\begin{equation}\label{alpha1}
	\begin{array}{c}
		\alpha(\bar{\chi},\bar{x})\triangleq
		\ln\left(1+\frac{||\bar{\chi}||^2}{\bar{x}}\right)-
\frac{||\bar{\chi}||^2}{\bar{x}},\\
		0<\psi(\bar{\chi},\bar{x})\triangleq \frac{||\bar{\chi}||^2}{\bar{x}(||\bar{\chi}||^2+\bar{x})}
	\end{array}
\end{equation}
and
\begin{equation}\label{ivt4}
	\frac{||\pmb{\chi}||^2}{\bx}\geq -\frac{||\bar{\chi}||^4}{\bar{x}^2}	+2\frac{\bar{x}+||\bar{\chi}||^2}{\bar{x}^2}\Re\{\bar{\chi}^H\pmb{\chi}\}-
\frac{||\bar{\chi}||^2}{\bar{x}^2}(||\pmb{\chi}||^2+\bx).
\end{equation}
The inequality (\ref{inv2}) has been proved in  \cite{TTN16}.  The following steps assist in
proving (\ref{ivt4}):
\begin{align}
	\frac{||\pmb{\chi}||^2}{\bx}&=\frac{1}{1-\frac{||\pmb{\chi}||^2}{\bx+||\pmb{\chi}||^2}}-1\label{ivt4a}\\
	&\geq \frac{2}{1-\frac{||\bar{\chi}||^2}{\bar{x}+|\bar{\chi}||^2}}-\frac{1}{\left(1-\frac{||\bar{\chi}||^2}{\bar{x}+||\bar{\chi}||^2}\right)^2}
	\left( 1-\frac{||\pmb{\chi}||^2}{\bx+||\pmb{\chi}||^2} \right)\nonumber\\&\quad-1\label{ivt4b}\\
	&=2\left(1+\frac{||\bar{\chi}||^2}{\bar{x}}\right)-\left(1+\frac{||\bar{\chi}||^2}{\bar{x}}\right)^2\nonumber\\
	&\quad+\left(1+\frac{||\bar{\chi}||^2}{\bar{x}}\right)^2\frac{||\pmb{\chi}||^2}{\bx+||\pmb{\chi}||^2}-1\label{ivt4c}\\
	&=-\frac{||\bar{\chi}||^4}{\bar{x}^2}+\frac{\left(|\bar{x}|+||\bar{\chi}||^2 \right)^2}{\bar{x}^2}\frac{||\pmb{\chi}||^2}{\bx+||\pmb{\chi}||^2}\label{ivt4d}\\
	&\geq-\frac{||\bar{\chi}||^4}{\bar{x}^2}+\frac{\left(|\bar{x}|+||\bar{\chi}||^2 \right)^2}{\bar{x}^2}\left(2\frac{\Re\{\bar{\chi}^H\pmb{\chi}\}}{\bar{x}+||\bar{\chi}||^2}\right.\nonumber\\
	&\left.\quad-\frac{||\bar{\chi}||^2}{\left(\bar{x}+||\bar{\chi}||^2 \right)^2}(||\pmb{\chi}||^2+\bx)\right) \label{ivt4e}\\
	&=-\frac{||\bar{\chi}||^4}{\bar{x}^2}
	+2\frac{\bar{x}+||\bar{\chi}||^2}{\bar{x}^2}\Re\{\bar{\chi}^H\pmb{\chi}\}-\frac{||\bar{\chi}||^2}{\bar{x}^2}(||\pmb{\chi}||^2+\bx),\label{ivt4f}
\end{align}
which is the same as (\ref{ivt4}). Both expressions (\ref{ivt4b}) and (\ref{ivt4e})
are derived through the linearization of the convex function $1/(1-\pmb{t})$ for $\pmb{t}=||\pmb{\chi}||^2/(\pmb{x}+||\pmb{\chi}||^2)\in [0,1)$.

The following inequality holds true for all $\pmb{\chi}_\ell\in\mathbb{C}^N$, $\bar{\chi}_\ell\in\mathbb{C}^N$,
and $\bx_\ell>0$, $\bar{x}_\ell>0$, $\ell\in\clL$, and $c>0$:
\begin{align}
	&\ln\left(\sum_{\ell\in\clL}(1-\frac{||\pmb{\chi}_\ell||^2}{||\pmb{\chi}_\ell||^2+c\bx_\ell})\right)\nonumber\\
	\leq&\ln\left(\sum_{\ell\in\clL}(1-\frac{||\bar{\chi}_\ell||^2}{||\bar{\chi}_\ell||^2+c\bar{x}_\ell})\right)\nonumber\\
	&+\left(\sum_{\ell\in\clL}(1-\frac{||\bar{\chi}_\ell||^2}{c\bar{x}_\ell+||\bar{\chi}_\ell||^2})\right)^{-1}\sum_{\ell\in\clL}\frac{||\bar{\chi}_\ell||^2}
	{c\bar{x}_\ell+||\bar{\chi}_\ell||^2}\nonumber\\
	&-\left(\sum_{\ell\in\clL}(1-\frac{||\bar{\chi}_\ell||^2}{c\bar{x}_\ell+||\bar{\chi}_\ell||^2})\right)^{-1}
	\sum_{\ell\in\clL}\left(2\frac{\Re\{\bar{\chi}_\ell^H\pmb{\chi}_\ell\}}
	{c\bar{x}_\ell+||\bar{\chi}_\ell||^2}\right.\nonumber\\  &\left.\quad-\frac{||\bar{\chi}_\ell||^2}{(c\bar{x}_\ell+||\bar{\chi}_\ell||^2)^2}(||\pmb{\chi}_\ell||^2+c\bx_\ell)\right).
	\label{ap6}
\end{align}
Define $\pi_\ell(\pmb{\chi}_{\ell},\bx_{\ell})\triangleq ||\pmb{\chi}_\ell||^2+c\bx_\ell$ to write
\begin{align}
&\ln\left(\sum_{\ell\in\clL}(1-\frac{||\pmb{\chi}_\ell||^2}{||\pmb{\chi}_\ell||^2+c\bx_\ell})\right)\nonumber\\
=&\ln\left(\sum_{\ell\in\clL}(1-\frac{||\pmb{\chi}_\ell||^2}{\pi_\ell(\pmb{\chi}_{\ell},\bx_{\ell})})\right)
\nonumber\\
\leq&\ln\left(\sum_{\ell\in\clL}(1-\frac{||\bar{\chi}_\ell||^2}
{\pi_\ell(\bar{\chi}_{\ell},\bar{x}_{\ell})})\right)+\left(\sum_{\ell\in\clL}(1-\frac{||\bar{\chi}_\ell||^2}
{\pi_\ell(\bar{\chi}_{\ell},\bar{x}_{\ell})})\right)^{-1}\nonumber\\
&\times\sum_{\ell\in\clL}\left(\frac{||\bar{\chi}_\ell||^2}
{\pi_\ell(\bar{\chi}_{\ell},\bar{x}_{\ell})}  -\frac{||\pmb{\chi}_\ell||^2}{\pi_\ell(\pmb{\chi}_{\ell},\bx_{\ell})}\right),\label{ap7}
\end{align}
while
\begin{equation}\label{ap8}
\frac{||\pmb{\chi}_\ell||^2}{\pi_\ell(\pmb{\chi}_{\ell},\bx_{\ell})}\geq
2\frac{\Re\{\bar{\chi}_\ell^H\pmb{\chi}_\ell\}-||\bar{\chi}_{\ell}||^2}
		{\pi_\ell(\bar{\chi}_{\ell},\bar{x}_{\ell})} -\frac{|\bar{\chi}_\ell||^2}{\pi^2_\ell(\bar{\chi}_{\ell},\bar{x}_{\ell})}
\pi_\ell(\pmb{\chi}_{\ell},\bx_{\ell}).
\end{equation}
Thus (\ref{ap6}) follows from (\ref{ap7}) and (\ref{ap8}).

Considering the both sides of (\ref{inv2}) and (\ref{ivt4}) ((\ref{ap6}), resp.) as the functions of $(\pmb{\chi},\pmb{x})$ ($(\pmb{\chi}_{\ell}, \pmb{x}_{\ell})$, $\ell\in\clL$, resp.),
the RHS of (\ref{inv2}) and (\ref{ivt4}) ((\ref{ap6}), resp.) match with their LHS counterpart at $(\bar{\chi},\bar{x})$  ($(\bar{\chi}_{\ell}, \bar{x}_{\ell})$, $\ell\in\clL$, resp.), so the latter is a tight minorant (majorant, resp.) of the former at
$(\bar{\chi},\bar{x})$ ($(\bar{\chi}_{\ell}, \bar{x}_{\ell})$, $\ell\in\clL$, resp.) \cite{Tuybook}.
\bibliographystyle{IEEEtran}
\balance \bibliography{mmwave}

\end{document}